\documentstyle[epsf,12pt]{article}

\begin{document}
\title{
\vspace{-40pt}
\hfill\parbox{3.5cm}{\normalsize
KUNS-1411\\HE(TH)~96/12\\
Niigata 96/E1\\
Aichi-5/96\\
patt-sol/9610009}\\
\vspace{15pt}
Solvable Optimal Velocity Models\\ and\\
Asymptotic Trajectory}
\author{Ken Nakanishi\thanks{e-mail address:
                nakanisi@gauge.scphys.kyoto-u.ac.jp}\\
                {\small\it Department of Physics,
                Kyoto University, Kyoto 606-01, Japan}
        \vspace{10pt} \\
        Katsumi Itoh\thanks{e-mail address:
                itoh@ed.niigata-u.ac.jp}\ \ \
        Yuji Igarashi\thanks{e-mail address:
                igarashi@ed.niigata-u.ac.jp}\\
                {\small\it Physics Division, Dept. of Education,
                Niigata University, Niigata 950-21, Japan}
        \vspace{10pt} \\
        Masako Bando\thanks{e-mail address:
                bando@aichi-u.ac.jp}\\
                {\small\it Physics Division,
                Aichi University, Miyoshi, Aichi 470-02, Japan}
        \vspace{10pt}
}
\date{October 23, 1996}
\maketitle
\vspace{-20pt}
\begin{abstract}
In the Optimal Velocity Model proposed as a new version of Car
Following Model, it has been found that a congested flow is generated
spontaneously from a homogeneous flow for a certain range of the
traffic density. A well-established congested flow obtained in a
numerical simulation shows a remarkable repetitive property such that
the velocity of a vehicle evolves exactly in the same way as that of
its preceding one except a time delay $T$. This leads to a global
pattern formation in time development of vehicles' motion, and gives
rise to a closed trajectory on $\Delta x$-$v$ (headway-velocity) plane
connecting congested and free flow points.  To obtain the closed
trajectory analytically, we propose a new approach to the pattern
formation, which makes it possible to reduce the coupled car following
equations to a single difference-differential equation (Rondo
equation).  To demonstrate our approach, we employ a class of linear
models which are exactly solvable. We also introduce the concept of
``asymptotic trajectory'' to determine $T$ and $v_B$ (the backward
velocity of the pattern), the global parameters associated with
vehicles' collective motion in a congested flow, in terms of
parameters such as the sensitivity $a$, which appeared in the original
coupled equations.
\end{abstract}

\section{Introduction}
\label{sec:Introduction}
Traffic flow is one of the most interesting phenomena of many-body
systems which may be controlled by basic dynamical equation.  Recent
developments in study of traffic flow has brought a renewed interest
in microscopic approaches, such as Optimal Velocity Model (OV-model)
\cite{AichiA,AichiB,AichiC}, which is a new version of Car Following
Model \cite{Pipes,Newell,Gazis}, Cellular Automaton Models
\cite{Biham,Nagel,Nagatani}, Coupled Map Lattice Models \cite{Kikuchi}
and Fluid Dynamical Models \cite{Kerner}.  The OV-model, among others,
has especially attracted interest because it provides us with a
possibility of unified understanding of both free and congested
traffic flows from common basic dynamical equations.  Unlike
traditional Car Following models, it introduces Optimal Velocity
Function $V(\Delta x)$ as a desirable velocity depending on headway
distance $\Delta x$.  The basic equation of OV-model for a series of
vehicles on a circuit of length $L$ is,
\begin{equation}
\ddot x_n(t) = a\{ V(\Delta x_n(t))-\dot x_n(t)\}\qquad n=1,2 \cdots N,
\label{eq:ovm0}
\end{equation}
where $x_n$ denotes the position of the $n$-th vehicle, $\Delta x_n
\equiv x_{n-1}-x_n$ headway and $N$ the total vehicle number.  The
constant parameter $a$ is the sensitivity.  A driver accelerates (or
decelerates) his vehicle in proportion to the difference between his
velocity and the optimal velocity $V(\Delta x)$.  As easily noticed, a
homogeneous flow is a solution to Eq.~(\ref{eq:ovm0}).  In such a flow,
vehicles have a common headway $L/N$, which is the inverse of the
vehicle density.  Stability of homogeneous flows is analyzed within a
linear approximation\cite{AichiA,AichiB}; it is stable for
$f=V'(L/N)<f_c$ and unstable for $f>f_c$.  The critical value is found
to be $f_c=a/2$.

In order to demonstrate that the OV-model describes ``spontaneous
generation of congestion'', numerical simulations were made using
Eq.~(\ref{eq:ovm0}). It was found that for $f<f_c$, i.e., if the
density is above the critical value, a slightly perturbed homogeneous
flow develops to a congested flow after enough time.  The congested
flow consists of alternating two distinct regions; congested regions
(high density), and smoothly moving regions or free regions (low
density). In this way the traffic congestion occurs spontaneously in
the OV-model.  This phenomenon can be understood as a sort of phase
transition from a homogeneous flow state to a congested flow state
\cite{AichiA,AichiB}.

A remarkable feature of the well-established congested flow is that
the velocity of the $n$-th vehicle ${\dot x_n}$ has the same time
dependence as that of the preceding ($(n-1)$-th) vehicle except a
certain time delay $T$.  It is also found that the global pattern
moves backward with a velocity $v_B$. This kind of behavior of
vehicles may be called ``repetitive pattern formation''.  It leads to
formation of a closed trajectory (``limit cycle'') on $\Delta x$-$v$
plane, along which representative points for all the vehicles move one
after another.  The convergence of vehicles' trajectories to a closed
trajectory signals the congestion in a traffic flow.  Therefore the
determination of the closed trajectory is one of the most important
subjects to understand congested flows. However this has been done
mostly in computer simulations.  The purpose of the present paper is
to obtain this closed trajectory directly by an analytical method.
Actually it has been found that, in the vicinity of the critical point
for the congested flow, equation (\ref{eq:ovm0}) can be reduced to the
modified KdV equation by the dynamical reduction method \cite{Sasa}.
In this paper we propose a new analytical approach to the pattern
formation, which may be applicable to any congested flow.

We argue that, once the repetitive pattern is formed, the coupled car
following equations reduce to a single difference-differential
equation (Rondo equation) for a universal function (Rondo function).
A Rondo function determines a closed trajectory on $\Delta x$-$v$
plane.  To make the Rondo equation tractable, we have simplified our
question on the following two points: firstly we have assumed that
OV-functions are piece-wise linear; secondly we have concentrated our
attention to an asymptotic trajectory, which is the key concept to be
explained in the next section.  We would like to stress that our
method does not lose its generality by making the above assumption on
OV-functions: an OV-function to be obtained from real data may be
approximated by a piece-wise linear function.

With the above simplifications, we have solved the Rondo equation for
each model and given an asymptotic trajectory on the $\Delta x$-$v$
plane.  Our result clearly tells us that, once an OV-function and the
sensitivity $a$ are given, an asymptotic trajectory is uniquely
determined; this then implies that the parameters $T$ and $v_B$ for a
collective motion of vehicles are given as a function of $a$.
Therefore our approach provides us with a method to determine $a$
dependence of the global parameters $T$ and $v_B$.

This paper is organized as follows.  Next section summarizes the main
results obtained from numerical simulations of the OV-model, with
emphasis on the pattern formation in a congested flow.  The concept of
an asymptotic trajectory is explained in this section.  As will become
clear in later sections, an asymptotic trajectory is a very important
concept to understand the OV-model.  We derive the
difference-differential equation and present a general strategy on how
to solve it in section~\ref{sec:Rondo Approach}.  This part summarizes
the central idea of this paper.  In order to demonstrate how the Rondo
approach works, we analytically solve, in section~\ref{sec:Piece-wise
Linear}, some simple models with piece-wise linear OV-functions.  Our
first model has been investigated by Sugiyama and
Yamada\cite{Sugiyama}.  Here we solve this model in the context of the
Rondo approach.  Although an asymptotic trajectory is very close to a
real trajectory observed in a computer simulation, it is not exactly
the same to the latter.  We describe some aspects of real trajectories
based on our knowledge of the asymptotic one in
section~\ref{sec:Trajectories around Cusps}.  The final section is
devoted to summary and discussions.

\section{Pattern Formation in OV-Model}
\label{sec:Pattern Formation}
Let us recollect what we have learned with numerical simulations of an
OV-model \cite{AichiA,AichiB,AichiC}. Suppose a simulation is
performed with a given OV-function and a fixed sensitivity $a$.  After
a congested flow is well-established, typical features of the
repetitive behavior can be observed in the following two figures.

\begin{figure}[hbt]
\hspace*{-0.5cm}
\epsfxsize=8.2cm
\epsfbox{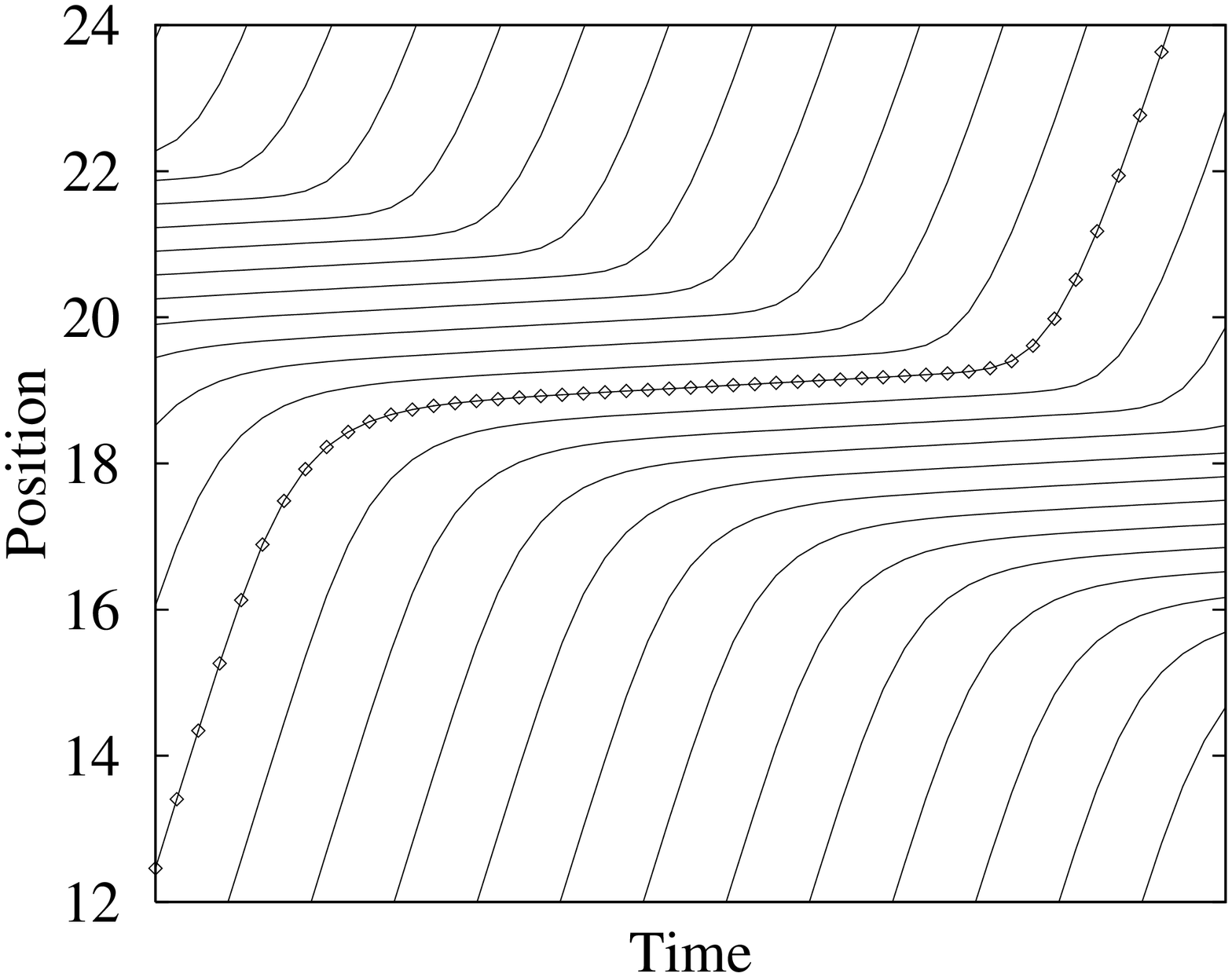}
\hspace*{0.3cm}
\epsfxsize=8.0cm
\epsfbox{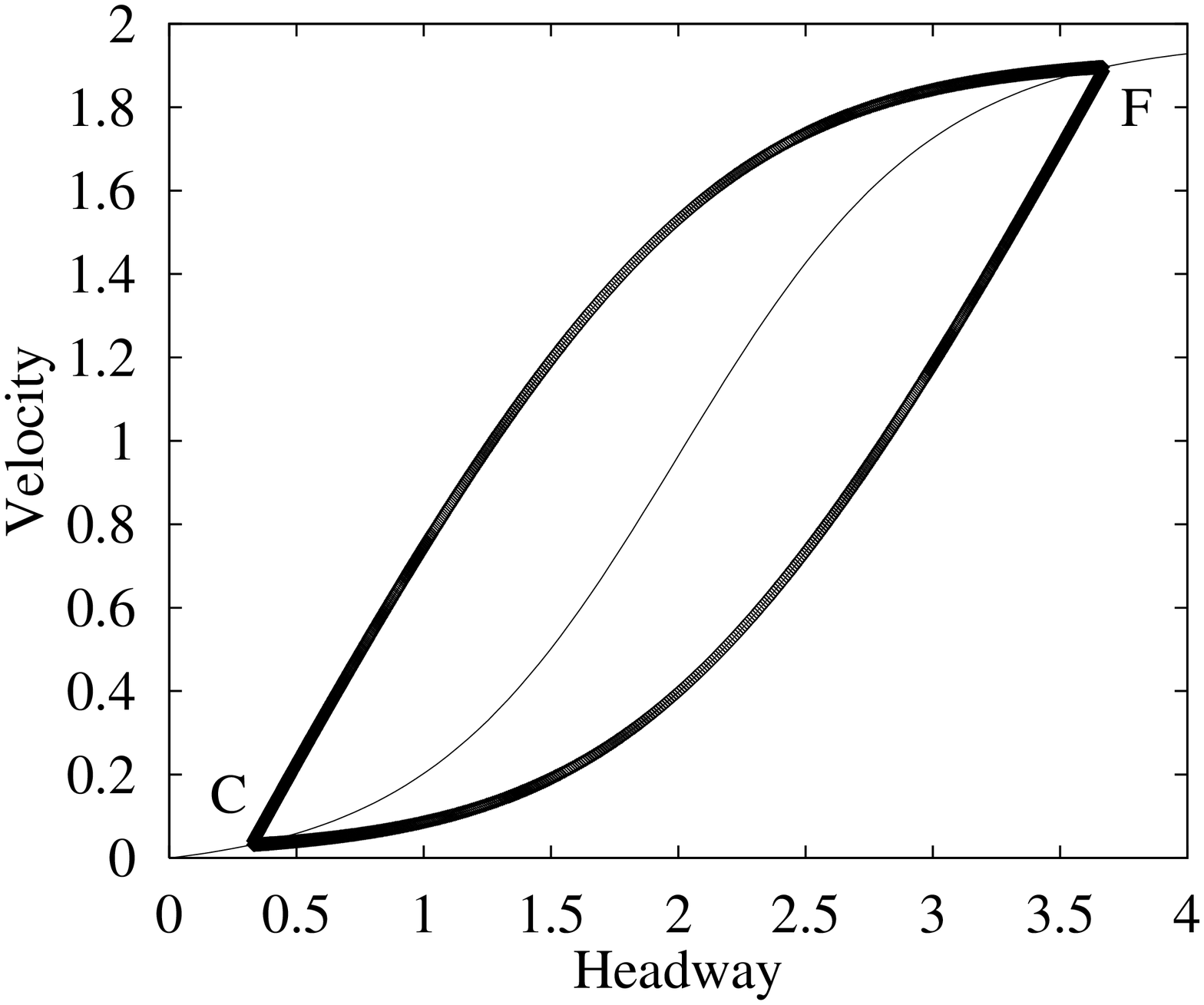}
\caption{
A result of the numerical simulation with 20 vehicles on a circuit
with the circumference $L=40$. The sensitivity and the OV-function are
chosen as $a=1.0$ and $V(h)=\tanh(h-2)+\tanh(2)$.  (a) Trajectories of
vehicles passing through a congested region.\ \ (b) The ``limit
cycle'' on the $\Delta x$-$v$ (headway-velocity) plane. The cusp
points C and F correspond to congested and free regions respectively.
Representative points for vehicles move anti-clockwise along this
``limit cycle''. They run fast on curves connecting the two cusps,
while they stay around cusps for a while. The thin line shows the
OV-function.}
\label{fig:intro.traje.eps}
\end{figure}

Figs.~\ref{fig:intro.traje.eps}(a) and (b) show that vehicles move in
alternating regions of free and congested flows.  It is recognized
that every vehicle behaves in the same manner as its preceding one
with a certain time delay $T$: as a result, the congested region moves
backward with the velocity $v_B$.  Once the location of any vehicle,
say, the $n$-th vehicle, is given as a function of $t$, we may
reproduce the pattern in Fig.~\ref{fig:intro.traje.eps}(a) by
plotting a series of functions shifted in time and position by $T$ and
$v_BT$ appropriately.  Therefore we expect that a congested flow may
be completely determined by a function of $t$ and global parameters
$T$ and $v_BT$.  The precise specifications of our approach to this
repetitive behavior will be explained in the next section.

Fig.~\ref{fig:intro.traje.eps}(b) clearly shows there exists a
``limit cycle'' on the $\Delta x$-$v$ plane, a closed curve with two
cusps at points C $(\Delta x_{\rm C},v_{\rm C})$ and F $(\Delta x_{\rm
F},v_{\rm F})$, both of which are on the OV-function.  At these cusp
points, we find $V'(\Delta x)<a/2$, which means that homogeneous flows
with such headways and velocities are linearly stable. Representative
points for all vehicles move on this ``limit cycle'' in the same
direction as anti-clockwise.  It follows from the conservation of
flow,
\begin{equation}
\frac{v_{\rm C}+v_B}{\Delta x_{\rm C}}=
\frac{v_{\rm F}+v_B}{\Delta x_{\rm F}}=\frac1T,
\label{eq:flow conserv}
\end{equation}
that the straight line connecting C and F has the slope $T^{-1}$ and
intersects with the vertical axis at $-v_B$.\footnote{This relation is
an approximate one except for an asymptotic trajectory to be discussed
below.}

\begin{figure}[hbt]
\epsfxsize=9cm
\centerline{
\epsfbox{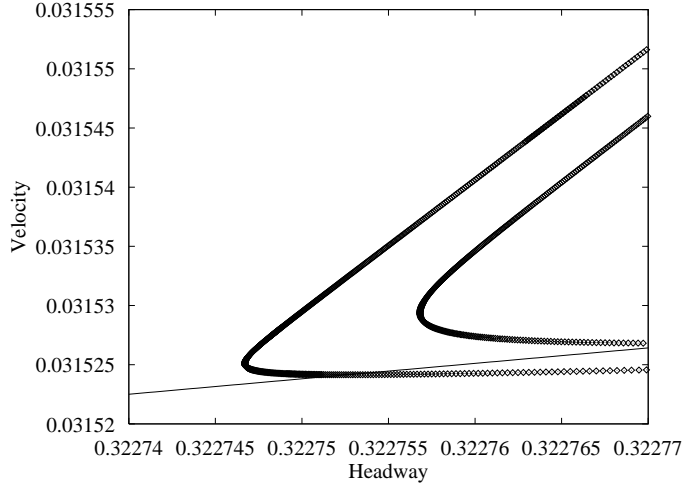}
}
\caption{
Trajectories for a vehicle passing through two congested regions
different in size. The outer curve corresponds to the larger congested
region. The thin line is the OV-function. Parameters are the same as
those in Figure 1.}
\label{fig:C-point.eps}
\end{figure}

In the rest of this section, we would like to explain the concept of
an asymptotic trajectory on the $\Delta x$-$v$ plane.  Suppose that a
vehicle passes through two congested regions different in size on a
circuit. Then the representative point moves on curves as shown in
Fig.~\ref{fig:C-point.eps}.  Each trajectory may not form an actual
cusp, rather it will form a round shaped tip. Also we find that the
larger the congested region, the sharper the shape of the tip: the
minimum velocity of the vehicle is smaller for a longer congestion.
It is rather easy to imagine that for a very short congestion the
vehicle cannot decelerate itself enough to reach the velocity
appropriate for a longer congested region.  If we plot minimum
velocities for longer and longer congested regions, we would find a
limiting value for the minimum velocities.  This value must be
realized for an infinitely long congested region.  With a similar
argument we find the limiting value for maximum velocities
corresponding to an infinitely long free region.  We may imagine the
following extreme situation: a vehicle starting from an infinitely
long free (congested) region goes toward an infinitely long congested
(free) region. The trajectory for this limiting situation will be
called as a decelerating (an accelerating) asymptotic trajectory.
Combining them we would find a closed curve with two real cusps on the
OV-function.

The duration for a vehicle to stay in a congested region would
obviously get longer for a larger congested region.  For an asymptotic
trajectory it becomes infinite. This does fit to our linear analysis
since the behavior of a vehicle is controlled by exponential functions
in time. In section~\ref{sec:Trajectories around Cusps}, we shall see
that exponential functions determine a curve near a cusp.

\section{Rondo Approach}
\label{sec:Rondo Approach}
We begin our description of the Rondo approach with two basic
assumptions:
\begin{enumerate}
\item  Velocities of the $n$-th and $(n-1)\,$-th vehicles have exactly
  the same time dependence if a certain time delay
  $T$ is taken into account, $\dot x_{n-1}(t) = \dot x_n(t+T)$.
\item The pattern of traffic flow moves backward with a constant
  velocity $v_B$.
\end{enumerate}
The above described properties are expressed as,
\begin{equation}
x_{n-1}(t) = x_n(t+T) + v_BT.
\label{eq:zentetsu2}
\end{equation}
All the vehicles' behavior is represented with a single universal
function $F(t)\equiv x_n(t)$;\footnote{This assumption bares some
similarities with the travelling wave ansatz ($\phi(x,t)\rightarrow
f(x-vt))$ for the wave equation.}
\begin{equation}
x_{n-k}(t) = F(t+kT) + kv_BT.
\label{eq:zentetsu3}
\end{equation}
With $n$-th vehicle's headway given as
\begin{equation}
\Delta x_n(t) = F(t+T) -F(t) + v_BT,
\label{eq:hw-v}
\end{equation}
$N$ coupled car following equations (\ref{eq:ovm0}) are reduced to a
single difference-differential equation for $F(t)$,
\begin{equation}
{1\over a}\ddot F(t) + \dot F(t) = V(F(t+T) -F(t) + v_BT).
\label{eq:zt-eq}
\end{equation}
In the following it will be called the Rondo equation.

In this paper, we will seek the Rondo function $F(t)$ for the
asymptotic trajectory.\footnote{We will discuss more realistic
situations with finite congested regions in section 5.} Before
studying concrete models, let us consider its generic properties.
Since the position and the velocity of vehicles are obviously
continuous in time, $F(t)$ is a continuously differentiable function.

An asymptotic trajectory connects the points F and C, each of which
corresponds to an infinitely long free or congested region (an
approximately homogeneous flow) satisfying the stability condition
mentioned in section~\ref{sec:Pattern Formation}.  Therefore $F(t)$
should be homogeneous flows asymptotically in the infinite past and
future: ${\dot F}(t)\rightarrow$ const. as $t\rightarrow \pm\infty$.
An asymptotic trajectory interpolates two stable solutions of the
equation (\ref{eq:ovm0}).  In this sense $F(t)$ may be regarded as a
``kink solution''.

Like OV-models studied in earlier papers, each model in the next
section has an OV-function which is symmetric with respect to a point,
S$(\Delta x_{\rm S},v_{\rm S})$.\footnote{The symmetry of an
OV-function is absolutely not necessary to solve a system in the Rondo
approach.  In the last section, we discuss how to solve the Rondo
equation for a generic situation.} So we assume this property in the
following arguments and quote our result in
appendix~\ref{ap-sec:symmetric}.  Two end points of an asymptotic
trajectory, C$(\Delta x_{\rm C},v_{\rm C})$ and F$(\Delta x_{\rm
F},v_{\rm F})$, are symmetric with respect to S.  Three points C, F
and S are on a straight line with a slope $T^{-1}$ and an intersection
$-v_B$.  Note that once the slope is given, the intersection is
uniquely determined since the point S must be on the line.

As shown in appendix~\ref{ap-sec:symmetric}, accelerating and
decelerating asymptotic trajectories are symmetric with respect to S.
Therefore it is sufficient to study one of them; in the rest of this
paper, we will take an decelerating asymptotic trajectory. Here we
summarize conditions which should be satisfied by the function $F(t)$:
\begin{enumerate}
\item[1)] $F(t)$ and $\dot F(t)$ are continuous for any $t$
($F(t)\in C^1$);
\item[2)] $v_{\rm F}=
\displaystyle\lim_{t\rightarrow -\infty}{\dot F}(t)$ and
$v_{\rm C}=\displaystyle\lim_{t\rightarrow +\infty}{\dot F}(t)$;
\item[3)] $v_{\rm F}+v_{\rm C}=2v_{\rm S}$ and
$\Delta x_{\rm F}+\Delta x_{\rm C}=2\Delta x_{\rm S}$, where
$\Delta x_{\rm F}=\displaystyle\lim_{t\rightarrow
-\infty}(F(t+T)-F(t)+v_BT)$ and $\Delta x_{\rm C}
=\displaystyle\lim_{t\rightarrow+\infty}(F(t+T)-F(t)+v_BT)$.
\end{enumerate}

We would like to explain a way to solve the Rondo equation, which
contains an OV-function and a sensitivity $a$ as well as $T$ and $v_B$
associated the pattern formation.  (1) First we give the parameter
$T$.  By drawing a straight line through the point S with the slope
$T^{-1}$, we find the intersection $-v_B$ and the points C and F.  (2)
Now $a$ is the only free parameter of the Rondo equation.  If we could
solve the equation, we would obtain a one-parameter family of (or
$a$-dependent) Rondo functions.  (3) Among this family, the right
Rondo function is selected by requiring that it connects the points C
and F.  This condition also determines a unique value for $a$.
Accordingly we find the $a$-dependence of $T$.

\section{Piece-wise Linear Function Models}
\label{sec:Piece-wise Linear}

We consider here a class of models with piece-wise linear
OV-functions. The Rondo equation is now linearized for all regions of
$\Delta x$, and therefore exactly solvable.

\subsection{Step Function Model}
The first model has the step function for the OV-function,
\begin{equation}
V(\Delta x)=\left\{
\begin{array}{ccl}
0\qquad&\Delta x < \Delta x_{\rm S}&({\rm region\ I}) \\
V_0\qquad&\Delta x > \Delta x_{\rm S}&({\rm region\ II}).
\end{array}\right.
\label{eq:SB}
\end{equation}
This OV-model has been solved in Ref.~\cite{Sugiyama}.  Here we explain
how this model can be solved in our Rondo approach.  In this model,
the Rondo equation is given by
\begin{equation}
\frac{1}{a}\ddot F(t)+\dot F(t)=
V_0\,\theta\left(F(t+T)-F(t)+v_BT-\Delta x_{\rm S}\right),
\end{equation}
where $\theta(x)$ is the Heviside function.

In the motion corresponding to a decelerating asymptotic trajectory,
the representative point for a vehicle moves from region~II into
region~I. Let us take the time $t$ such that the point moves into the
region~I at $t=0$, which implies $\Delta x_n(0)=\Delta x_{\rm S}$.

The equation of motion is
\begin{equation}
\frac{1}{a}\ddot F(t)+\dot F(t)=\left\{
\begin{array}{cl}
V_0\qquad&(t < 0)\\
0\qquad&(t > 0).
\end{array}\right.
\label{eq:SA}
\end{equation}
The general solutions for two regions are
\begin{equation}
\dot F(t)=\left\{
\begin{array}{ll}
V_0 + C_1 e^{-at}\qquad& (t \le 0)\\
C_2 e^{-at}\qquad& (t \ge 0).
\end{array}\right.
\label{eq:SA1}
\end{equation}
The integration constants are determined as $C_1=0$ and $C_2=V_0$ from
requirements that ${\dot F}(t)$ be a continuous function and
asymptotically constant for $t \rightarrow \pm \infty$.  The
continuous function $F(t)$ is
\begin{equation}
F(t)=\left\{
\begin{array}{ll}
V_0 t\qquad& (t \le 0)\\
{\displaystyle\frac{V_0}{a}}(1-e^{-at})\qquad& (t \ge 0).
\end{array}\right.
\label{eq:SA2}
\end{equation}
Here we choose the origin of position coordinate so that $F(0)=0$.

The function $F(t)$ with its relation to the headway, $\Delta x_n(t) =
F(t+T) - F(t) + v_B T$, completely determines a decelerating
asymptotic trajectory on the $\Delta x$-$v$ plane. From the
condition~3) in the section~\ref{sec:Rondo Approach}, the asymptotic
trajectory connects symmetric points on the OV-function. This gives us
a condition $\Delta x_n(+\infty) + \Delta x_n(-\infty) = 2 \Delta
x_{\rm S} = 2 \Delta x_n(0)$,
\begin{equation}
v_BT+(V_0+v_B)T=2\left[\frac{V_0}{a} (1 -  e^{-aT})+v_BT\right].
\label{eq:SI}
\end{equation}
This may be expressed as the transcendental equation for $\rho=aT$
\begin{equation}
e^{-\rho} + \frac{1}{2}\rho - 1 = 0.
\label{eq:SJ}
\end{equation}
Since the $\rho$ is found to be the constant (1.59362..), we obtain
\begin{equation}
 aT = \rho=1.59362..\,.
\label{eq:SK}
\end{equation}
This gives us the $a$-dependence of $T$, which was first obtained in
Ref.~\cite{Sugiyama}.

\begin{figure}[htb]
\hspace*{-0.5cm}
\epsfxsize=8cm
\epsfbox{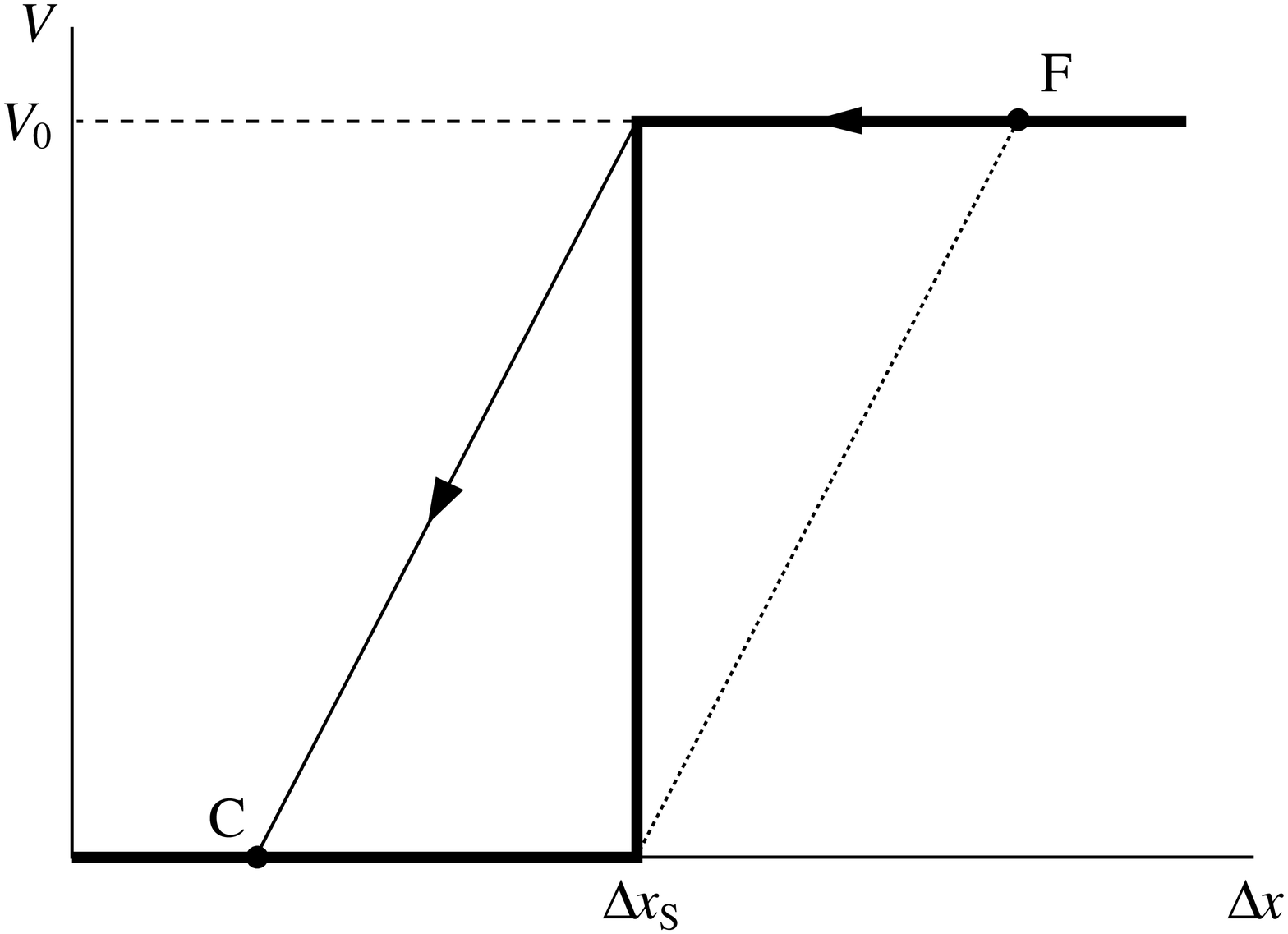}
\hspace*{0.5cm}
\epsfxsize=8cm
\epsfbox{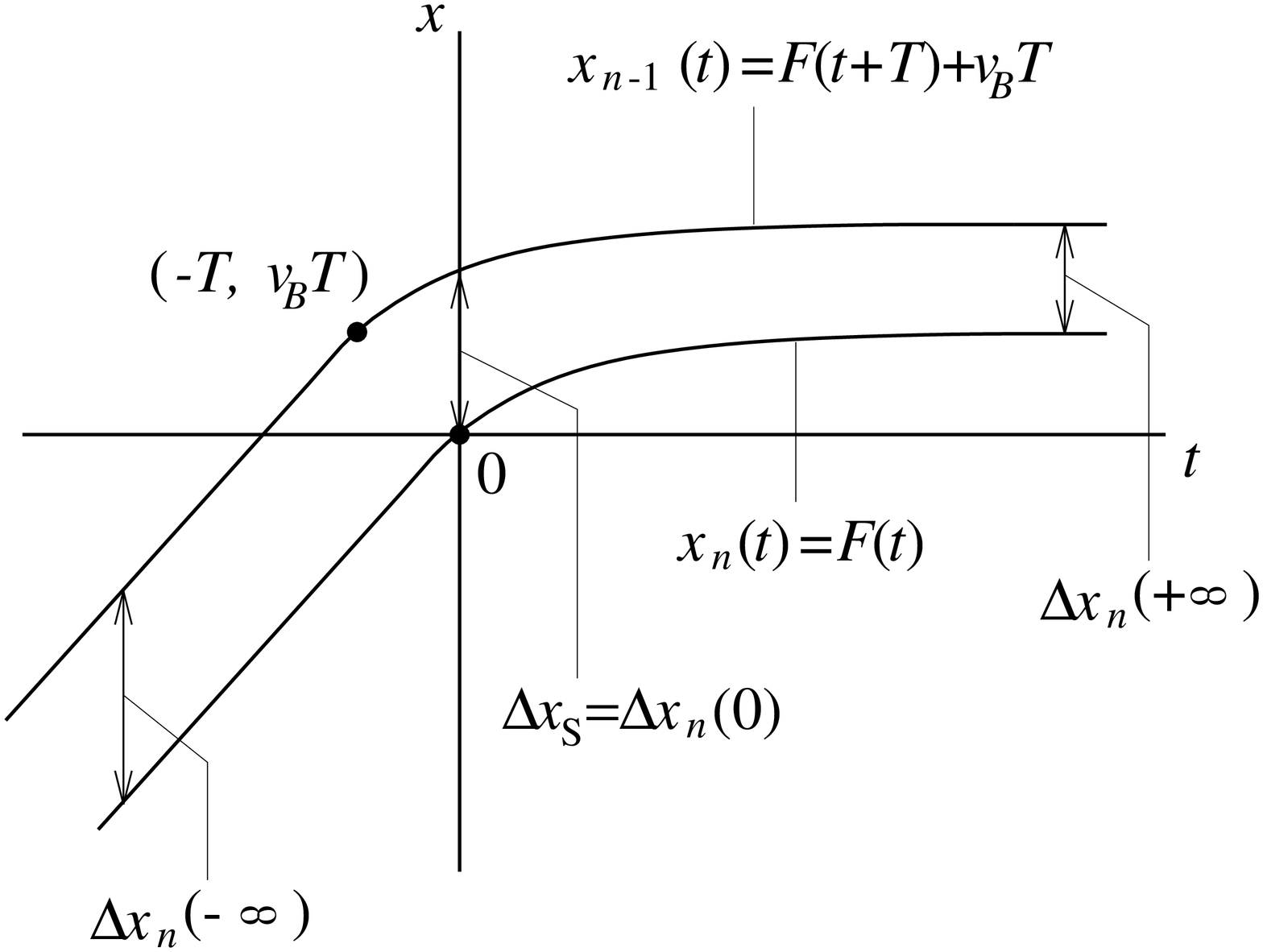}
\caption{
(a) The OV-function (thick line) and a decelerating asymptotic
trajectory (thin line with arrows) for the Step Function Model.\ \ (b)
The position functions $x_n(t)$ and $x_{n-1}(t)$ for the decelerating
asymptotic trajectory.}
\label{fig:step.eps}
\end{figure}
It is instructive to see a relation between the function $F(t)$ and
the decelerating trajectory depicted in Fig.~\ref{fig:step.eps}.  Two
curves in Fig.~\ref{fig:step.eps}(b) correspond to the $(n-1)$-th and
$n$-th vehicles' locations.  At $t=0$ the $n$-th vehicle's
representative point moves into the region~I and $F(t)$ is described
by an exponential function. Before that time, the function $F(t)$ is
linear in $t$.  The curve for the $(n-1)$-th vehicle changes from the
linear to the exponential behavior at $t=-T$.  It is given via a
parallel displacement by the vector $(-T, v_BT)$ from the curve
$F(t)$.  For the time $t \le -T$, curves are two parallel straight
lines, which implies the headway of the $n$-th vehicle does not change
till that time from the infinite past.  The $n$-th vehicle has the
constant velocity $V_0$ for $t < 0$.  This tells us that the $n$-th
vehicle is in the free region for $t \le -T$, indicated by the point F
in Fig.~\ref{fig:step.eps}(a).  It is also easy to observe that at
$t=-T$ the $(n-1)$-th vehicle starts to decelerate.  As a result the
headway of the $n$-th vehicle decreases; at $t=0$ it reaches the value
$\Delta x_{\rm S}$ and the $n$-th vehicle starts to decelerate itself.

In this model, the points F and C are both characterized as points
which are reached in the infinite future or past.  As a traffic flow,
we are describing a soliton like solution connecting a half infinite
vehicles, running with the velocity $v_F$ and the headway $\Delta
x_F$, and another half infinite vehicles going into the congested
region associated with the point C.

\subsection{Single Slope Function Model}

\begin{figure}[hbt]
\epsfxsize=8cm
\centerline{
\epsfbox{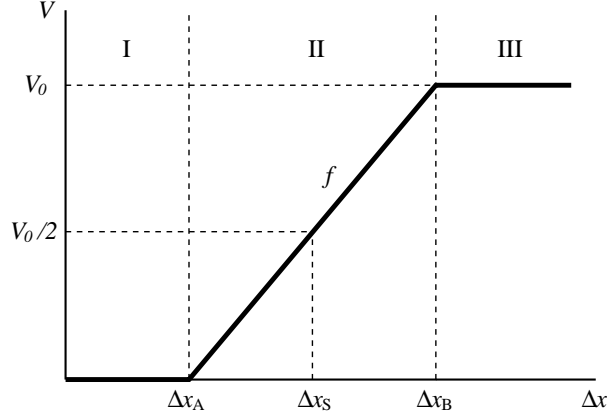}
}
\caption{
The OV-function for the Single Slope Function Model. $f$ is the
gradient in the region~II. The function is symmetric with respect to
the point $(\Delta x_{\rm S},V_0/2)$.}
\label{fig:piOVF.eps}
\end{figure}

In the step function model, the $\Delta x$-dependence of the Rondo
equation is too simple; $T$-dependence is not explicit.  So we would
like to consider a slightly improved model, whose OV-function shown in
Fig.~\ref{fig:piOVF.eps} has a finite slope.  Thus we call this model
as the Single Slope Function Model.  The OV-function is characterized
by the following parameters: $f$, the slope; $V_0$, the maximum
velocity; $\Delta x_{\rm S}$, the headway for optimal velocity
$V_0/2$.
{}From the linear analysis in the paper \cite{AichiA}, we know that a
homogeneous flow becomes unstable and a congested flow is expected for
$2f>a$. This condition is assumed in our analysis here.

The OV-function has sharp bends at
\begin{eqnarray}
\Delta x_{\rm A}=\Delta x_{\rm S}-\frac{V_0}{2f},\\
\Delta x_{\rm B}=\Delta x_{\rm S}+\frac{V_0}{2f},
\label{eq:delta x_A}
\end{eqnarray}
which divide $\Delta x$ into three regions, I, II and III, as
indicated in Fig.~\ref{fig:piOVF.eps}.

We would like to find a Rondo function $F(t)$ for a decelerating
asymptotic trajectory, along which headway of a vehicle monotonically
decreases from $\Delta x_{\rm F}$ to $\Delta x_{\rm C}$. We assume it
reaches $\Delta x_{\rm B}$ at $t=-\tau$ and $\Delta x_{\rm A}$ at
$t=0$. The Rondo equation takes of the form,
\begin{equation}
\frac{1}{a}{\ddot F}(t)+{\dot F}(t)=\left\{
\begin{array}{ll}
\, 0\qquad& (t \ge 0)\\
f\left(F(t+T)-F(t)-\delta\right)\qquad& (-\tau \le t \le 0)\\
V_0\qquad& (t \le -\tau)
\end{array}
\right.,
\end{equation}
subject to the conditions,
\begin{eqnarray}
F(T)-F(0)+v_BT
&\kern-0.2cm=&\kern-0.2cm\Delta x_{\rm A},\label{eq:t=0}\\
F(-\tau+T)-F(-\tau)+v_BT&\kern-0.2cm=&\kern-0.2cm
\Delta x_{\rm B}.\label{eq:t=-tau}
\end{eqnarray}
Here $\delta= \Delta x_A - v_B T$.  Note that the time $-\tau$ is not
a free parameter.  Rather, it should be determined from
(\ref{eq:t=-tau}) via solving the Rondo equation.

For regions~I and III, the Rondo equation becomes the same as that in
the step function model.  Therefore, we obtain
\begin{equation}
\dot F(t)=\left\{
\begin{array}{ll}
{\rm const.} \times e^{-at}\qquad& ({\rm for\ region~I})\\
\,V_0\qquad& ({\rm for\ region~III}).
\end{array}\right.
\label{eq:SAb}
\end{equation}
Our purpose in this subsection is to describe a method to find the
Rondo function in the region~II which correctly interpolates those in
(\ref{eq:SAb}).  To this end, we may introduce a series of functions
for each time interval:
\begin{equation}
F(t)=\left\{
\begin{array}{ll}
F^{\rm I}(t)\qquad& (0 \le t)\\
F_1(t)\qquad& (-T \le t \le 0)\\
F_2(t)\qquad& (-2T \le t \le -T)\\
\quad\vdots& \qquad\vdots\\
F^{\rm III}(t)\qquad& (t \le -\tau)
\end{array}
\right.
\end{equation}
The first function $F^{\rm I}(t)\equiv F_0(t)$ and the last one
$F^{\rm III}(t)$ are for the region~I and~III, respectively.

Let us find $F^{\rm I}(t)$ ( for $t \ge 0$ ).  It follows from
(\ref{eq:SAb}) that
\begin{eqnarray}
{\dot F}^{\rm I}(t)&\kern-0.2cm=&\kern-0.2cm u_0 e^{-at},\\
F^{\rm I}(t)
&\kern-0.2cm=&\kern-0.2cm\frac{u_0}{a}\left(1-e^{-at}\right),
\end{eqnarray}
where we fix again $F^{\rm I}(0)=0$. The $n$-th vehicle's headway is
then given by
\begin{equation}
\Delta x_n(t)=F^{\rm I}(t+T)+v_B T-F^{\rm I}(t)
=\frac{u_0}{a}(1- e^{-aT})e^{-at}+v_B T.
\end{equation}
Since the condition (\ref{eq:t=0}) determines the constant $u_0$
\begin{equation}
u_0=a\frac{\Delta x_{\rm A}-v_B T}{1- e^{-aT}}
\equiv\frac{a\delta}{1- e^{-aT}},
\end{equation}
we find the Rondo function
for $t \ge 0$ to be
\begin{equation}
F^{\rm I}(t)=\frac{\delta}{1- e^{-aT}}\left(1-e^{-at}\right).
\label{eq:FI}
\end{equation}
This leads to a linear relation between the velocity and the headway,
\begin{equation}
{\dot x_n} =  \frac{a}{1-e^{-aT}} (\Delta x_n - v_B T).
\end{equation}
In the $t \rightarrow \infty$ limit, we find that ${\dot x_n}
\rightarrow 0$ and $\Delta x_n \rightarrow v_B T$.

Now we consider $F_1(t)$ ( for $-T \le t \le 0$ ).
In this time interval, the Rondo equation for $F_1(t)$ is expressed as
\begin{equation}
\frac 1a{\ddot F_1}(t) + {\dot F_1(t)}
= f ( F_0(t+T)-F_1(t)- \delta).
\end{equation}
in terms of the Rondo function for the region~I, $F^{\rm I}(t)\equiv
F_0(t)$.  By using the differential operator,
\begin{equation}
{\cal D}=\frac 1{af}\frac{d^2}{dt^2}+\frac 1f\frac{d}{dt}+1,
\end{equation}
we may rewrite the above equation as
\begin{equation}
{\cal D}F_1(t)= F_0(t+T)-\delta.
\label{eq:F1}
\end{equation}
The general solution may be written as a sum of a particular solution
and the solution to the homogeneous equation ${\cal D}F_1^{\rm
hom}(t)=0$.  It is easy to see that $F_0(t+T)-\delta$ is a particular
solution, since the function $F_0(t)\equiv F^{\rm I}(t)$ given in
(\ref{eq:FI}) satisfies
\begin{equation}
{\cal D}F_0(t)= F_0(t).
\label{eq:F0}
\end{equation}
The exponents $\gamma$ for a homogeneous solution are
\begin{equation}
\gamma=-\frac a2 \pm i \sqrt{af-\frac{a^2}4}
\equiv -\frac a2 \pm i\omega.
\label{eq:omega}
\end{equation}
A general solution is given as
\begin{equation}
F_1(t) = F_0(t+T)-\delta
+ e^{-\frac a2t}\left(A\sin{\omega t}+B\cos{\omega t} \right).
\end{equation}
The constants $A$ and $B$ are determined as $A=a\delta/\omega$ and
$B=0$, by the requirement that $F_0(t)$ and $F_1(t)$ must be
continuous up to the first derivative at $t=0$.  Therefore
\begin{equation}
F_1(t)=\frac{\delta}{1- e^{-aT}}\left(1-e^{-a(t+T)}\right)
-\delta+\frac{a\delta}\omega e^{-\frac a2t}\sin{\omega t},
\label{eq:sol-F1}
\end{equation}
and the headway and the velocity for the $n$-th vehicle is given as
\begin{eqnarray}
\Delta x_n(t)&\kern-0.2cm=&\kern-0.2cm
F_0(t+T)-F_1(t)+v_BT=\Delta x_{\rm A}
-\frac{a\delta}\omega e^{-\frac a2t}\sin{\omega t},\cr
\dot x_n(t)&\kern-0.2cm=&\kern-0.2cm
\dot F_1(t)=\frac{a\delta}{e^{aT}-1}e^{-at}
+\frac{a\delta}\omega e^{-\frac a2t}
\left(\omega\cos{\omega t}-\frac{a}2\sin{\omega t}\right).
\label{eq:h&v-F1}
\end{eqnarray}

We would now like to give general formula for $F_k(t)$ ( for $t\in
I_k=[-kT,-(k-1)T]$ ) with $k>1$.  Suppose that $F_k(t)$ for $t\in I_k$
is known to us and we are trying to find $F_{k+1}(t)$ for $t\in
I_{k+1}$. $F_{k+1}(t)$ satisfies the second order linear differential
equation,
\begin{equation}
{\cal D}F_{k+1}(t)= F_k(t+T)-\delta,
\label{eq:Fk+1}
\end{equation}
with boundary conditions;
\begin{eqnarray}
F_{k+1}(-kT)&\kern-0.2cm=&\kern-0.2cmF_k(-kT),\cr
{\dot F}_{k+1}(-kT)&\kern-0.2cm=&\kern-0.2cm{\dot F}_k(-kT),
\label{eq:connect-cond}
\end{eqnarray}
while $F_k(t)$ satisfies a similar equation,
\begin{equation}
{\cal D}F_k(t)= F_{k-1}(t+T)-\delta.
\label{eq:Fk}
\end{equation}

The function $F_k(t)$ describes the behavior of the $n$-th vehicle
only for $t \in I_k$.  However we will find it useful to define the
function by the relation (\ref{eq:Fk}) even outside the interval
$I_k$.  The function used outside the interval will be denoted as
${\widetilde F}_k(t)$.  The difference of (\ref{eq:Fk+1}) and
(\ref{eq:Fk}) gives us the following equation for $t\in
I_{k+1}=[-(k+1)T, -kT]$,
\begin{equation}
{\cal D}\left(F_{k+1}(t)-{\widetilde F}_k(t)\right)
=F_k(t+T)-{\widetilde F}_{k-1}(t+T)\quad (k\ge 1).
\label{eq:diff-eq}
\end{equation}
{}From (\ref{eq:F1}) and (\ref{eq:F0}) we find
\begin{equation}
{\cal D}\left(F_1(t)-{\widetilde F}_0(t)\right)
=F_0(t+T)-{\widetilde F}_0(t)-\delta =\delta\left(e^{-at}-1\right),
\label{eq:diff-eq-k0}
\end{equation}
for $t\in I_1=[-T,0]$.

By using the function $G_k(t)$ defined in the relation,
\begin{equation}
F_{k+1}(t)={\widetilde F}_k(t)+G_{k+1}(t+kT),
\end{equation}
(\ref{eq:diff-eq}) and (\ref{eq:diff-eq-k0}) are rewritten into the
following equations for $-T\le t\le 0$,
\begin{eqnarray}
&&{\cal D}G_{k+1}(t)=G_k(t)\qquad (k\ge 1),
\label{eq:Gk}\\
&&{\cal D}G_1(t)=\delta\left(e^{-at}-1\right)\equiv G_0(t).
\label{eq:G1}
\end{eqnarray}
Note that $G_0(t)$ defined in (\ref{eq:G1}) satisfies ${\cal D}G_0(t)
=G_0(t)$.  The conditions (\ref{eq:connect-cond}) become $G_k(0)=0$
and ${\dot G}_k(0)=0$ ( for $k\ge1$ ).

The Rondo function $F(t)$ for the region~II is given as a sum of
$G_k(t)$,
\begin{equation}
F^{\rm II}(t)=F_0(t)+\sum_{k \ge 0} \theta(-t-kT)G_{k+1}(t+kT),
\label{eq:FII}
\end{equation}
for $t > -\tau$.  In the appendix~\ref{ap-sec:General formula} we will
give general solutions to differential equations (\ref{eq:Gk}) and
(\ref{eq:G1}).  $G_1(t)$ from the appendix,
\begin{equation}
G_1(t)=G_0(t)+\frac {a\delta}\omega e^{-\frac a2t}\sin{\omega t}=
\delta\left(e^{-at}-1\right)+
\frac {a\delta}\omega e^{-\frac a2t}\sin{\omega t},
\end{equation}
is consistent to (\ref{eq:sol-F1}).  Similarly $G_2(t)$ is given as
\begin{equation}
G_2(t)=G_1(t)-\frac {a\delta}\omega \frac{2f}{4f-a}
e^{-\frac a2t}\left\{\omega t\cos{\omega t}-\sin{\omega t}\right\}.
\end{equation}
So $F_2(t)$ for $t\in I_2=[-2T,-T]$ becomes
\begin{eqnarray}
F_2(t)&\kern-0.2cm=&\kern-0.2cm{\widetilde F}_1(t)+G_2(t+T)\cr
&\kern-0.2cm=&\kern-0.2cm{\widetilde F}_1(t)
+\delta\left(e^{-a(t+T)}-1\right)
+\frac {a\delta}\omega e^{-\frac a2(t+T)}\sin{\omega(t+T)}\cr
&&-\ \frac {a\delta}\omega \frac{2f}{4f-a}e^{-\frac a2(t+T)}
\left\{\omega(t+T)\cos{\omega(t+T)}-\sin{\omega(t+T)}\right\},
\end{eqnarray}
while the headway for the $n$-th vehicle is
\begin{eqnarray}
\Delta x_n(t)&\kern-0.2cm=&\kern-0.2cm\Delta x_{\rm A}-
\frac {a\delta}\omega e^{-\frac a2t}\sin{\omega t}\cr
&&+\ \frac {a\delta}\omega \frac{2f}{4f-a}e^{-\frac a2(t+T)}
\left\{\omega(t+T)\cos{\omega(t+T)}-\sin{\omega(t+T)}\right\}.
\end{eqnarray}
The general formula in appendix~\ref{ap-sec:General formula} may be
used further to generate $F_3,F_4,\cdots$, needed to describe the
trajectory in the region~II.

Let us consider the region~III. At $t=-\tau$, the function
(\ref{eq:FII}) for the region~II must be continuously connected to
$F^{\rm III}(t)$ including their first derivatives.  This condition
yields $F^{\rm III}(t)$ for $t\le-\tau$ as
\begin{equation}
F^{\rm III}(t)=V_0\ (t+\tau)+F^{\rm II}(-\tau).
\end{equation}
The continuity condition for the first derivative requires also that
\begin{equation}
\dot F^{\rm II}(-\tau)=\dot F^{\rm III}(-\tau)\equiv V_0.
\label{eq:dot-FII}
\end{equation}
This completes our construction of an asymptotic trajectory.

\begin{figure}[htb]
\epsfxsize=8cm
\centerline{
\epsfbox{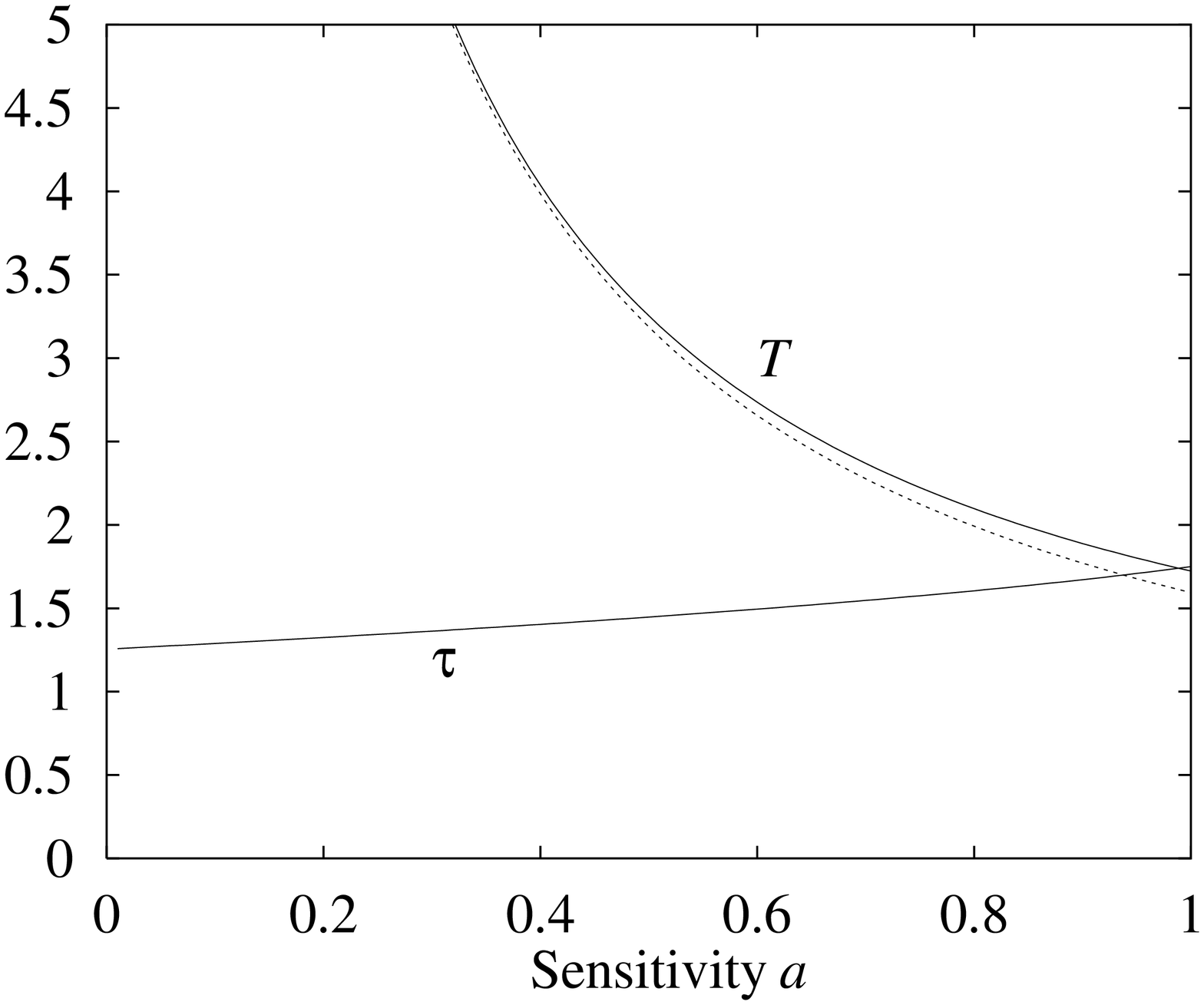}
}
\caption{
Sensitivity dependence of $T$ and $\tau$ for the Single Slope Function
Model.  The condition, $T\ge\tau$, is assumed, which is satisfied for
$a\le0.98857..$\,. The dotted line shows the $a$-$T$ relation for the
Step Function Model.}
\label{fig:aT.eps}
\end{figure}
Now let us find the $a$-dependence of $T$ for the present model.  The
time $-\tau$ may be determined by the condition (\ref{eq:t=-tau}),
$F(-\tau+T)-F^{\rm II}(-\tau)+v_BT=\Delta x_B$; then
Eq.~(\ref{eq:dot-FII}) gives us a relation between $a$ and $T$.  By
using (\ref{eq:FII}) for $F^{\rm II}(-\tau)$, concrete expressions of
(\ref{eq:t=-tau}) and (\ref{eq:dot-FII}) may be obtained.  For $-T \le
-\tau$, $F^{\rm II}(-\tau)$ is simply given by $F_1(-\tau)$ and the
above conditions are expressed as:
\begin{eqnarray}
&&\frac{a\delta}\omega e^{\frac a2\tau}\sin{\omega\tau}=
\Delta x_{\rm B}-\Delta x_{\rm A}\equiv\frac{V_0}{f},\\
&&\frac{a\delta}{e^{aT}-1} e^{a\tau}
+\frac{a\delta}\omega e^{\frac a2\tau}
\left(\omega\cos{\omega\tau}+\frac{a}2\sin{\omega\tau}\right)=V_0.
\end{eqnarray}
The requirement of symmetry, $\Delta x_n(+\infty)+\Delta x_n(-\infty)=
2\Delta x_{\rm S}$, mentioned in section~\ref{sec:Rondo Approach}
becomes $v_BT+(V_0+v_B)T=2\Delta x_{\rm S}$.  This relation and
Eq.~(\ref{eq:delta x_A}) allow us to express $\delta=\Delta x_A - v_B
T$ as,
\begin{equation}
\delta=(fT-1)\frac{V_0}{2f}.
\label{eq:delta}
\end{equation}
Finally, we reach to the coupled equations which determine $T$ and
$\tau$ for given slope $f$ and sensitivity $a$:
\begin{equation}
\left\{\begin{array}{l}
(fT-1) e^{\frac a2\tau}\sin{\omega\tau}
=\displaystyle\frac{2\omega}a,\\
\left(e^{aT}-1\right)
\left[\left(f-\displaystyle\frac{a}2\right)\sin{\omega\tau}
-\omega\cos{\omega\tau}\right]=\omega e^{\frac a2\tau},\end{array}
\right.
\label{eq:a-T}
\end{equation}
where $\omega$ is given in Eq.~(\ref{eq:omega}).  For $f=1.0$,
Eq.~(\ref{eq:a-T}) is solved numerically to give the $a$-dependence of
$T$ and $\tau$ as shown in Fig.~\ref{fig:aT.eps}. Since we used
$F_1(t)$ for $F^{II}(t)$, Eq.~(\ref{eq:a-T}) is valid only for $\tau
\le T$ ($\tau$ coincides with $T$ when $a=0.98857..$ at
$\tau=T=1.74027..$).

We observe in Fig.~\ref{fig:aT.eps} that $T$ behaves as $1/a$ for
small $a$: this implies that, for congested flows to be formed, the
delay $T$ must be larger for less sensitive drivers.  In the limit of
$f\rightarrow\infty$, the present model reduces to the Step Function
Model, in which $aT=\rho$ and $\tau=0$.  This may be confirmed with
Eq.~(\ref{eq:a-T}) since $aT$ reaches a constant, $\rho =1.59362..$,
and $a \tau$ behaves like $2a/(f\rho)$ when $a/f$ goes to
zero.\footnote{Note that Eq.~(\ref{eq:a-T}) may be rewritten in terms
of rescaled variables $aT$, $a \tau$ and $a/f$.}

\begin{figure}[htb]
\hspace*{-0.5cm}
\epsfxsize=8cm
\epsfbox{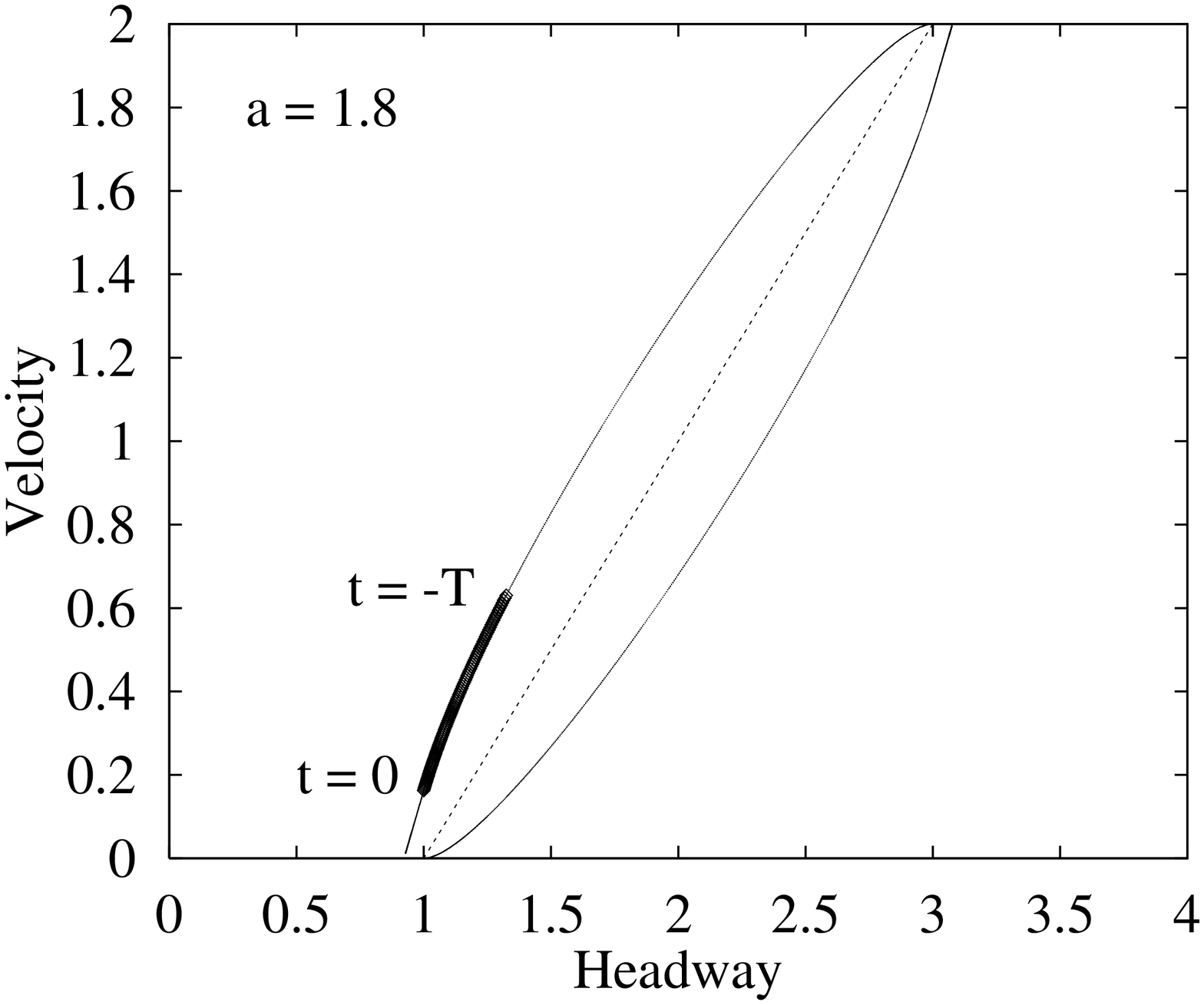}
\hspace*{0.5cm}
\epsfxsize=8cm
\epsfbox{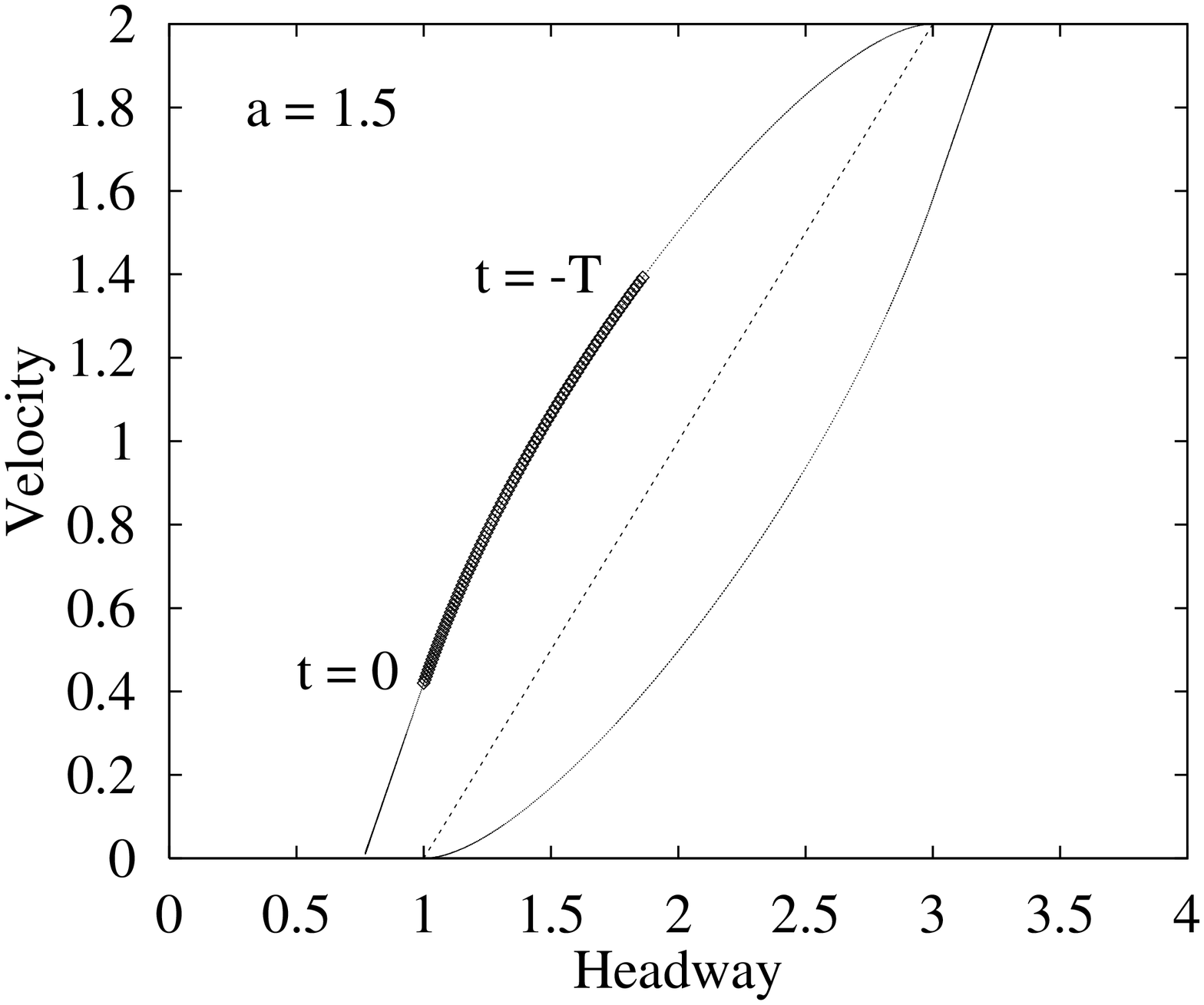}
\\
\hspace*{-0.5cm}
\epsfxsize=8cm
\epsfbox{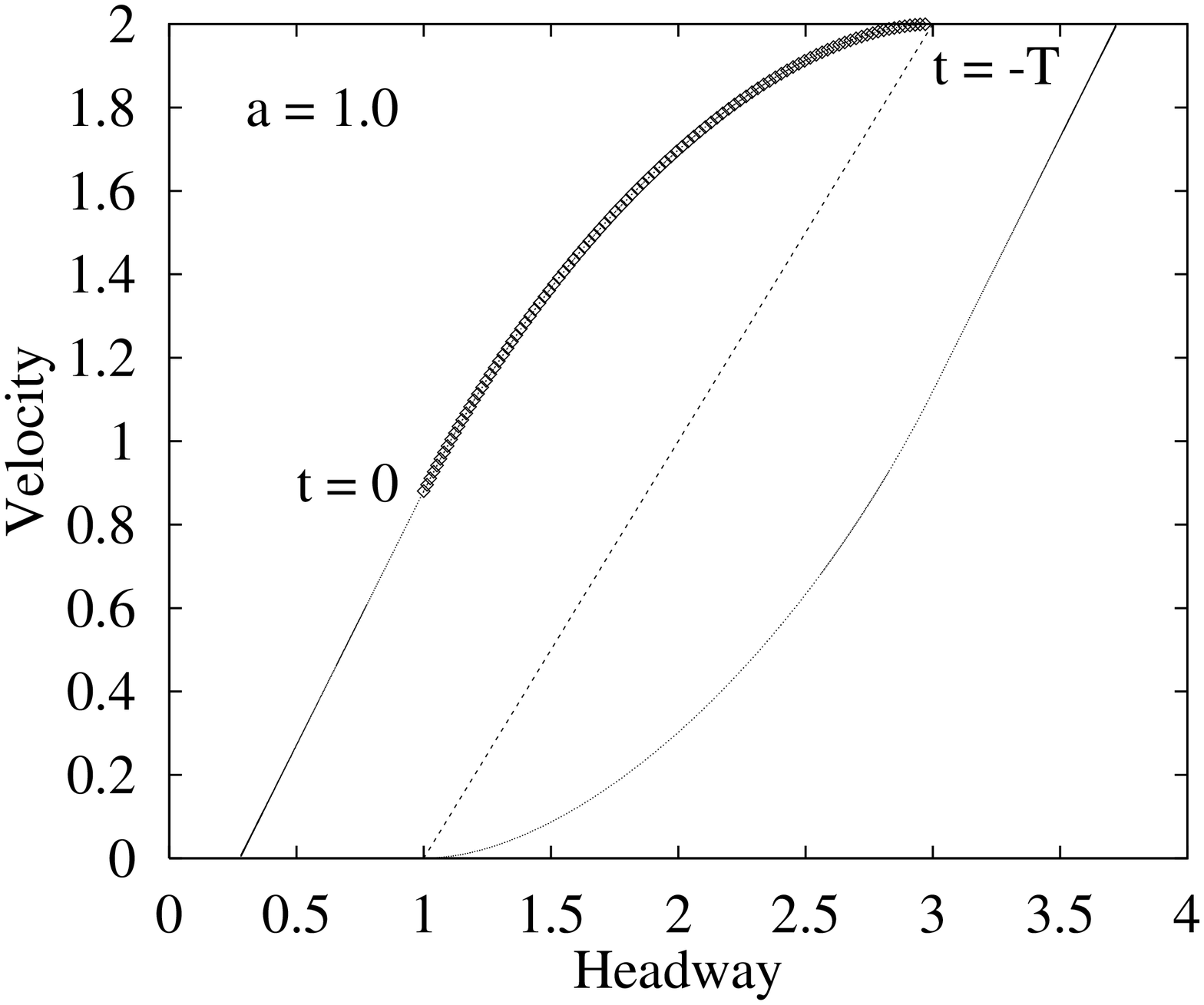}
\hspace*{0.5cm}
\epsfxsize=8cm
\epsfbox{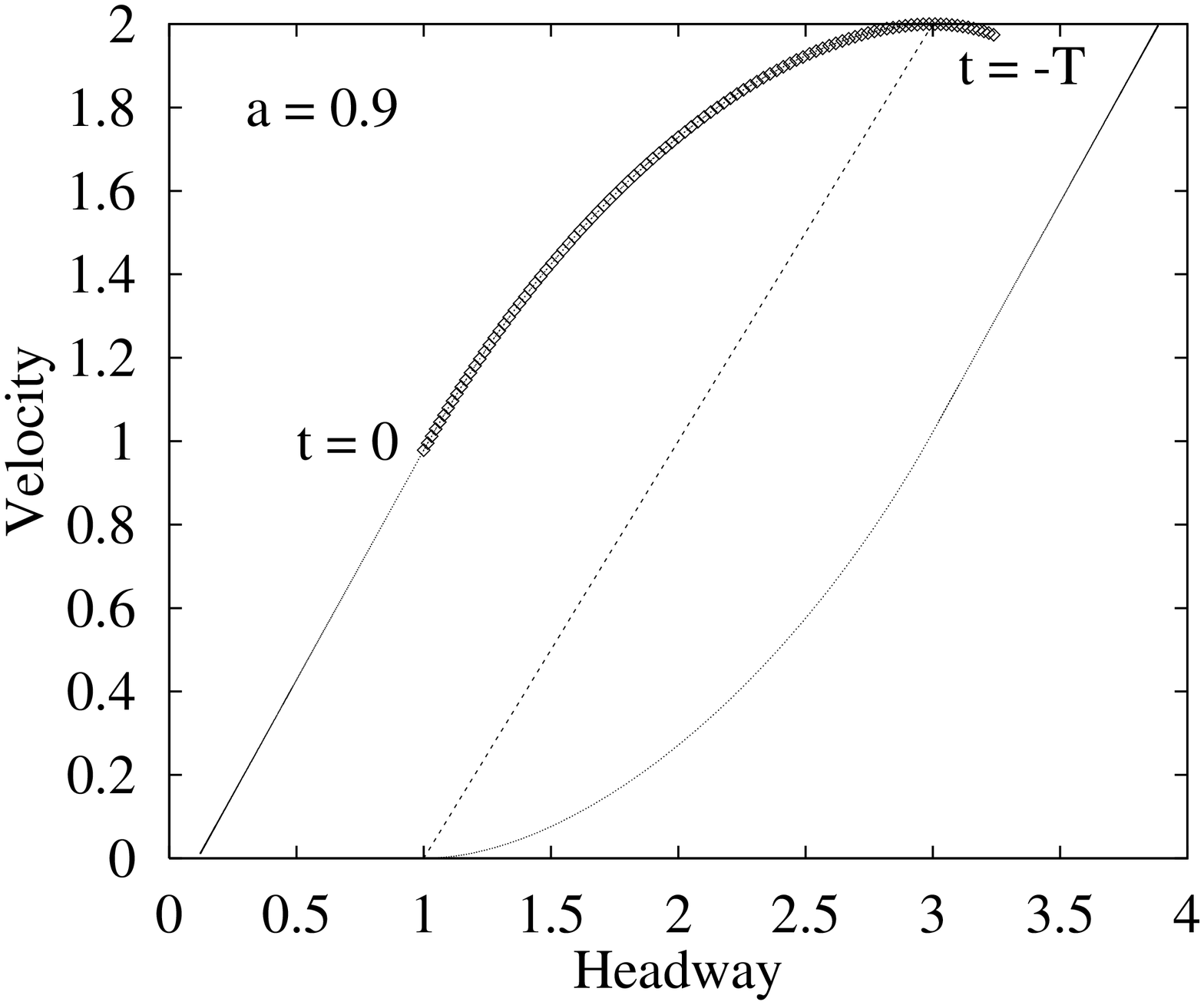}
\caption{
Thin curves show ``limit cycles'' for the Single Slope Function Model
obtained from simulations with (a)~$a=1.8$, (b)~$a=1.5$, (c)~$a=1.0$
and (d)~$a=0.9$.  Trajectories generated from $F_1(t)$ for $t \in
[-T,0]$ are drawn with thick curves. The curves inside the region~II
($1< \Delta x < 3$) are parts of asymptotic trajectories.  The dotted
line shows the OV-function with $f=1.0$, $V_0=2.0$ and $\Delta x_{\rm
S}=2.0$.}
\label{fig:1-slope.eps}
\end{figure}
In order to see the validity of our approach, let us compare our
results and trajectories obtained by simulations. In
Fig.~\ref{fig:1-slope.eps}, thick curves show parts of the
asymptotic trajectories, to be determined by the function $F_1(t)$,
for $a=1.8,\ 1.5\ (\tau>T)$, $a=1.0\ (\tau\sim T)$ and $a=0.9\
(\tau<T)$.  These curves are to be compared with numerical simulations
shown as thin curves.  The function $F_1(t)$ is enough to give an
asymptotic trajectory for $a\le0.98857..$, as mentioned above.  It is
expected that when $a$ gets closer to its critical value ($a_{\rm
critical}=2f=2.0$, in this case), we need functions $F_k(t)$ with
higher $k$ to form an entire trajectory.

There appears a flat trajectory again in the region~III.  As in the
Step Function Model, it takes time $T$ for a vehicle to move on the
flat trajectory and there is only one vehicle traveling on this
interval.

\subsection{Double Slope Function Model}
The OV-function for the single slope model has flat regions~I and~III
like the step function model.  For those regions the Rondo equation
does not depend on $F(t+T)$: a vehicle does not react to the motion of
the preceding one.  This motivates us to consider a more realistic
model with an OV-function which has non-zero gradient for any headway.

\begin{figure}[htb]
\epsfxsize=8cm
\centerline{
\epsfbox{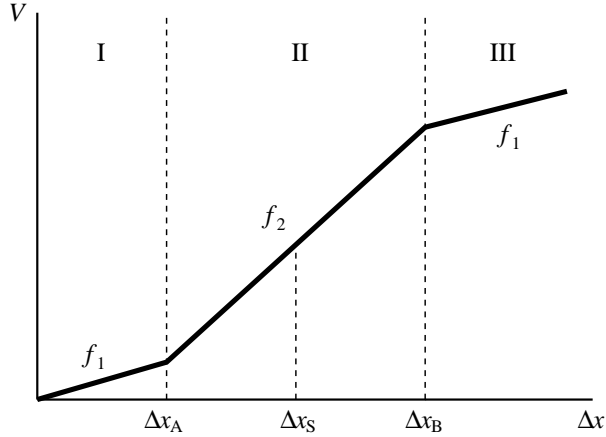}
}
\caption{
The OV-function for the Double Slope Function Model, which is
symmetric with respect to the point $(\Delta x_{\rm S}, V(\Delta
x_{\rm S}))$.  It has the gradient $f_1$~($f_2$) in the regions~I and
III (II).}
\label{fig:wOVF.eps}
\end{figure}
The OV-function has a slope
$f_1$ in regions~I and III, and $f_2$ in region~II
(see Fig.~\ref{fig:wOVF.eps}):
\begin{equation}
V(\Delta x)=\left\{
\begin{array}{lrl}
f_1\Delta x & \Delta x \le \Delta x_{\rm A}&{\rm (region\ I)}\\
f_2\left[\Delta x - (1-\kappa)\Delta x_{\rm A}\right] &
\Delta x_{\rm A}\le\Delta x\le\Delta x_{\rm B}&{\rm (region\ II)}\\
f_1\left[\Delta x+(\kappa^{-1}-1)
(\Delta x_{\rm B}-\Delta x_{\rm A})\right] &
\Delta x_{\rm B} \le \Delta x &{\rm (region\ III)}.
\end{array}\right.
\end{equation}
where $\kappa=f_1/f_2$. Obviously this function is symmetric around
$(\Delta x_{\rm S}, V(\Delta x_{\rm S}))$, where $\Delta x_{\rm S}$ is
the middle point of the region~II.  Here we should note that the
sensitivity $a$ must satisfy $f_1<a/2<f_2$ for generation of the
congestion in this model, since the homogeneous flow is expected to be
linearly unstable only in the region~II.

As in the previous subsection, we assume that the headway reaches
$\Delta x_{\rm B}$ at $t=-\tau$ and $\Delta x_{\rm A}$ at $t=0$. The
solution of the Rondo equation (\ref{eq:zt-eq}) must satisfy the
conditions (\ref{eq:t=0}), (\ref{eq:t=-tau}). The Rondo function
$F(t)$ for three regions will be denoted as follows;
\begin{equation}
F(t)=\left\{
\begin{array}{ll}
F^{\rm I}(t)\qquad & (0 \le t)\\
F^{\rm II}(t)\qquad & (-\tau \le t \le 0)\\
F^{\rm III}(t)\qquad & (t \le -\tau).
\end{array}
\right.
\end{equation}

For $F^{\rm I}(t)$ ($t \ge 0$), the Rondo equation is given by
\begin{equation}
{\cal D}_1F^{\rm I}(t)= F^{\rm I}(t+T)+v_B T,
\label{eq:DF1}
\end{equation}
where
\begin{equation}
{\cal D}_1=
\frac 1{af_{1}}\frac{d^2}{dt^2}+\frac{1}{f_1}\frac{d}{dt}+1.
\end{equation}
To find a solution to the homogeneous equation ${\cal D}_{1}F^{\rm
I}_{\rm hom}(t)= F^{\rm I}_{\rm hom}(t+T)$, we use the ansatz $F^{\rm
I}_{\rm hom}(t)\sim e^{\gamma t}$ which gives an equation for the
exponent $\gamma$,
\begin{equation}
\frac{\gamma^2}{a}+\gamma=f_{1}(e^{\gamma T}-1).
\label{eq:gamma}
\end{equation}
As long as the condition $f_1<1/T$ holds,\footnote{When a congested
flow is realized in this model, this condition is satisfied
trivially.} there are two real solutions: the negative $-{\gamma}_{\rm
in}$ and the positive one ${\gamma}_{\rm out}$.  Because of the
asymptotic behavior, $\dot F(t)\rightarrow v_{\rm C}$ as
$t\rightarrow+\infty$, only the exponent $-{\gamma}_{\rm in}$ is
relevant to the function $F^{\rm I}(t)$.
Adding a particular solution of (\ref{eq:DF1}), we obtain the solution
subject to conditions (\ref{eq:t=0}) and $F(0)=0$ as
\begin{equation}
F^{\rm I}(t)=v_{\rm C}t+\left(\Delta x_{\rm A}-\Delta x_{\rm C}\right)
\frac{1 - e^{-{\gamma}_{\rm in}t}}{1- e^{-{\gamma}_{\rm in} T}}.
\label{eq:FF0}
\end{equation}

By calculating $\Delta x_n(t)$ and $\dot x_n(t)$ from (\ref{eq:FF0}),
\begin{eqnarray}
\Delta x_n(t)&\kern-0.2cm=&\kern-0.2cm\Delta x_{\rm C}+
(\Delta x_{\rm A}-\Delta x_{\rm C})e^{-\gamma_{\rm in}t},\label{eq:Dx0}\\
\dot x_n(t)&\kern-0.2cm=&\kern-0.2cm v_{\rm C}+
(\Delta x_{\rm A}-\Delta x_{\rm C})\frac{\gamma_{\rm in}}
{1-e^{-\gamma_{\rm in}T}}e^{-\gamma_{\rm in}t},
\end{eqnarray}
we obtain a linear trajectory given by
\begin{equation}
\dot x_n(t)-v_{\rm C}=\frac{\gamma_{\rm in}}{1-e^{-\gamma_{\rm in}T}}
\left(\Delta x_n(t)-\Delta x_{\rm C}\right).
\end{equation}

In the region~II the function $F^{\rm II}(t)$ is divided into
$F_k(t)$'s for $t\in I_k=[-kT,-(k-1)T]$ as was done in the previous
subsection.  We may now study the Rondo equation,
\begin{equation}
{\cal D}_2F_k(t)= F_{k-1}(t+T)+v_BT-(1-\kappa)\Delta x_{\rm A}
\qquad (k \ge 1),
\label{eq:DF2}
\end{equation}
where $F_0(t)\equiv F^{\rm I}(t)$ and ${\cal D}_2$ is ${\cal D}_1$
with $f_1$ replaced by $f_2$: ${\cal D}_2={\cal
D}_1-(1-\kappa)\left({\cal D}_1-1\right)$. In terms of ${\cal D}_2$,
equation~(\ref{eq:DF1}) becomes
\begin{equation}
{\cal D}_2F_0(t)=F_0(t+T)+v_BT-(1-\kappa)\Delta x_n(t),
\end{equation}
where $\Delta x_n(t)$ is given by (\ref{eq:Dx0}).

The basic technique in the last subsection may be used to solve
(\ref{eq:DF2}) with a slight modification.  Let us define $G_k(t)$'s
with $F_{k+1}(t)={\tilde F}_k(t)+G_{k+1}(t+kT)$, which satisfy the
equations,
\begin{eqnarray}
&&{\cal D}_2G_{k+1}(t)=G_k(t)\qquad (k\ge 1),
\label{eq:GGk}\\
&&{\cal D}_2G_1(t)=\delta\left(e^{- \gamma_{\rm in}t}-1\right)
\equiv G_0(t),
\label{eq:GG1}
\end{eqnarray}
where $\delta=(1-\kappa)(\Delta x_{\rm A}-\Delta x_{\rm C})$.  The
boundary conditions are $G_k(0)={\dot G}_k(0)=0$ for $k \ge 1$.  In
solving these, we may use again the formula given in the
appendix~\ref{ap-sec:General formula}. Once we find the series of
$G_k(t)$, we obtain the function $F^{\rm II}(t)$ by the relation
(\ref{eq:FII}).  Further we can determine the time $-\tau$ by the
condition (\ref{eq:t=-tau}), $F(-\tau+T)-F^{\rm II}(-\tau)+v_BT=\Delta
x_B$.

In the region~III ($t\le-\tau$), the Rondo equation take the form,
\begin{equation}
{\cal D}_1F^{\rm III}(t)=F^{\rm X}(t+T)+v_BT+
(\kappa^{-1}-1)(\Delta x_{\rm B}-\Delta x_{\rm A}),
\label{eq:FIII}
\end{equation}
where $F^{\rm X}(t+T)$ is $F^{\rm II}(t+T)$ for $-\tau\le t+T\le -
\tau+T$ and $F^{\rm III}(t+T)$ for $t+T \le -\tau$. The solution must be
continuously connected to $F^{\rm II}(t)$ including the first derivative
at $t=-\tau$. Eq.~(\ref{eq:FIII}) can be solved by the same manner as in
the region~II, though homogeneous solutions to the equation
(\ref{eq:FIII}) include the hyperbolic functions instead of the
trigonometric functions.

The logic at the end of section 3 may be used to get a better
perspective on what we have discussed up to now, and it will leads us
to find the $a$-dependence of $T$.  First, with a given slope
$T^{-1}$, we draw the straight line through the point S.  The value of
$-v_B$ and the coordinates of C and F are given in terms of $T$ and
the parameters in the OV-function,
\begin{equation}
v_{\rm C}=\frac{f_1T}{1-f_1 T}v_B,\quad
\Delta x_{\rm C}=\frac{v_BT}{1-f_1 T},
\label{eq:h&v_C}
\end{equation}
\begin{eqnarray}
v_{\rm F}=
\frac{f_1v_BT+f_1(\kappa^{-1}-1)(\Delta x_{\rm B}-\Delta x_{\rm A})}
{1-f_1 T},\cr
\Delta x_{\rm F}=
\frac{v_BT+f_1T(\kappa^{-1}-1)(\Delta x_{\rm B}-\Delta x_{\rm A})}
{1-f_1 T},
\label{eq:h&v_F}
\end{eqnarray}
and
\begin{equation}
-v_B = \left( f_1 - \frac{f_2T+1}{2T} \right) \Delta x_A +
       \frac{f_2T-1}{2T} \Delta x_B.
\label{eq:vB}
\end{equation}

In solving the Rondo equation, we have introduced a parameter $\tau$
determined by Eq.~(\ref{eq:t=-tau}).  Using this $\tau$ and $v_B$
expressed as Eq.~(\ref{eq:vB}), we obtain one parameter family of
($a$-dependent) solution to the Rondo equation.  Then, we find an
appropriate value of $a$ for a given $T$ by the requirement that the
asymptotic trajectory connects C and F: $\dot F^{\rm
III}(-\infty)=v_{\rm F}$.
\begin{figure}[hbt]
\epsfxsize=10cm
\centerline{
\epsfbox{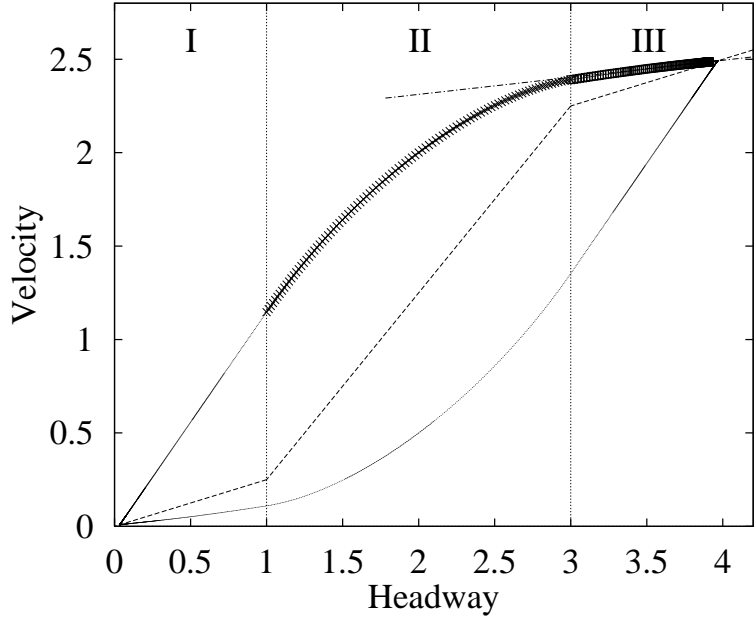}
}
\caption{
Trajectories for the Double Slope Model. Thin curve shows ``limit
cycle'' from simulation with $a=1.13124$ when $T=\tau=1.58331$. The
analytical solution is drawn with thick curve.  The dotted-dashed line
shows the trajectory expressed by the exponential function with index
$\gamma_{\rm out}$. The dotted line shows the OV-function with
$f_1=0.25$, $f_2=1.0$, $\Delta x_{\rm A}=1.0$ and $\Delta x_{\rm
B}=3.0$.}
\label{fig:w.eps}
\end{figure}

In Fig.~\ref{fig:w.eps}, we show trajectories from our analytic
study and a computer simulation for the double slope model.  We have
chosen a particular value for $a$ so that $\tau=T$ and we used the
Rondo function for $t>-\tau-T$ to draw the part of the decelerating
asymptotic trajectory.  Clearly the Rondo function reproduces the
trajectory obtained via a computer simulation.

By the procedure described in this subsection, we may easily obtain the
remaining part of the asymptotic trajectory.  When this is carried out,
we expect that it reaches to the point F in the infinite past.  In the
following we will give another argument to support this expectation.
The Rondo equation for $t < -\tau -T$ in the region~III may be solved
with exponential functions plus a particular solution, as for the
region~I: the function $F^{\rm III}(t)$ is the sum of a term linear in
$t$ and exponential functions.  In the limit $t \rightarrow - \infty$,
only the linear term survives expressing that vehicles have the velocity
for a free region, $v_{\rm F}$; the exponents satisfy
Eq.~(\ref{eq:gamma}) and have positive real parts so that exponential
functions vanish as $t \rightarrow -\infty$.

It would be appropriate to explain how solutions to
Eq.~(\ref{eq:gamma}) are distributed on the complex $\gamma$-plane.
Without going into the details, here we only quote features relevant
to our arguments.  There is only one solution with negative real part,
it actually a real solution $-\gamma_{\rm in}$.  Other solutions have
positive real parts, which are relevant when $t \sim -\infty$.  There
is only one real solution, $\gamma_{\rm out}$; there are complex pair
solutions with their real parts larger than $\gamma_{\rm out}$.  Since
$\gamma_{\rm out}$ has the smallest positive real part, it dominates
among exponential functions when $t \rightarrow -\infty$.  Thus for
very large negative $t$, the function $F^{\rm III}(t)$ may be
approximated by the exponential functions with $\gamma_{\rm out}$.  In
the region~I, we found that the decelerating asymptotic trajectory is
linear on the $\Delta x$-$v$ plane owing to the exponential term in
$F^{\rm I}(t)$.  Similarly, the approximated $F^{\rm III}(t)$ define a
line on the $\Delta x$-$v$ plane, starting the point F as shown in
Fig.~\ref{fig:w.eps}.  We observe that this line is actually the
tangent line to the trajectory at the point F.

The absence of the flat region is directly related to our observation
that the point F in reached only in the infinite past.  If we consider
the limit to have a flat region, $f_1=0$, we only have a negative
solution $\gamma_{\rm in}=-a$.  When we would like to consider the
OV-model applied for a realistic situation, the relevant OV-function
may be approximately realized by a piece-wise linear function.  Since
it is unlikely that the function has a flat region, the above feature
of the double slope model must be generic.

\section{Trajectories around Cusps}
\label{sec:Trajectories around Cusps}
In the last section, we discussed asymptotic trajectories in models
with piece-wise linear OV-functions.  The asymptotic trajectories can
be realized only when the number of vehicles becomes infinite. In
computer simulations in Refs.\cite{AichiA,AichiB,AichiC}, a finite
number of vehicles run around a circuit; a vehicle goes through all
the free and congested regions in a finite time.  The Rondo equation
probably has solutions even for such situations.  Though we have not
worked out how to obtain entire trajectories for such vehicles, we are
able to discuss parts of the trajectories around the points C or F.
We are going to discuss this subject in this section.

To make our explanation concrete, we take a trajectory around the end
point C. In a congested region, all the vehicles have almost the same
velocity. When a vehicle is about to reach a congested region, its
velocity would be slightly different from $v_{\rm C}$ and the function
$F(t)$ may be written as $F(t)=v_{\rm C} t+ \xi(t)$.  We take a linear
approximation for an OV-function $V(\Delta x)$\footnote{Here as an
OV-function we have in mind a smooth, but not necessarily a piece-wise
linear, function.} around the point ($\Delta x_{\rm C},v_{\rm C}$),
\begin{equation}
V(\Delta x)\simeq f_{\rm C}(\Delta x - \Delta x_{\rm C}) + v_{\rm C}.
\end{equation}
Since $(v_B + v_{\rm C})T=\Delta x_{\rm C}$, the Rondo equation
(\ref{eq:zt-eq}) becomes a linear difference-differential equation for
$\xi(t)$,
\begin{equation}
\frac{\ddot \xi(t)}{a} + {\dot \xi(t)}
= f_{\rm C} (  \xi(t+T) - \xi(t)).
\label{eq:linear_eq}
\end{equation}
For the ansatz $\xi(t)= e^{\gamma t}$, we find an equation for the
exponent $\gamma$,
\begin{equation}
\frac{\gamma^2}{a}+\gamma=f_{\rm C}(e^{\gamma T}-1).
\end{equation}
As long as $f_{\rm C}<1/T$, there are two real solutions: the
negative, $-{\gamma}_{\rm in}$, and the positive one, ${\gamma}_{\rm
out}$.\footnote{We ignored complex solutions in this approximation
since the real of those solutions are larger than ${\gamma}_{\rm
out}$.  See the discussion in the last section.}

\begin{figure}[hbt]
\epsfxsize=8cm
\centerline{
\epsfbox{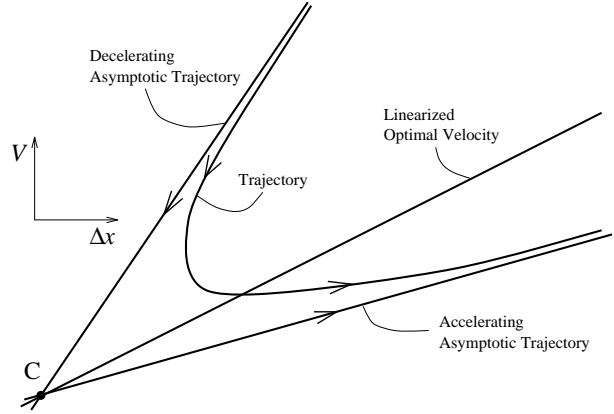}
}
\caption{
A trajectory in linear approximation for a vehicle passing through the
congested region with a finite size. It has two asymptotes,
accelerating and decelerating asymptotic trajectories.  The time
development is indicated with arrows.}
\label{fig:linearized.eps}
\end{figure}

Trajectories considered here cross (at $t=0$) the OV-function at
points slightly different from C: we denote their coordinates by
$(\Delta x,v)=(\Delta x_{\rm C}+\delta v/f_{\rm C},v_{\rm C}+\delta
v)$.  We find the solution for $F(t)$ as,
\begin{equation}
F(t)=v_{\rm C} t +
\frac{-{\gamma}_{\rm out}^2 e^{-{\gamma}_{\rm in}t}  +
{\gamma}_{\rm in}^2 e^{{\gamma}_{\rm out}t}} {{\gamma}_{\rm
in}{\gamma}_{\rm out}({\gamma}_{\rm in}+{\gamma}_{\rm out})}\delta v.
\end{equation}
We may find $\Delta x_n(t)$ and $v_n(t)$ from this solution
\begin{eqnarray}
\Delta x_n(t)&\kern-0.2cm=&\kern-0.2cm\Delta x_{\rm C} +
\frac{(a-{\gamma}_{\rm in}){\gamma}_{\rm out} e^{-{\gamma}_{\rm in}t}+
(a+{\gamma}_{\rm out}){\gamma}_{\rm in} e^{{\gamma}_{\rm out}t}}
{a({\gamma}_{\rm in}+{\gamma}_{\rm out})}{\delta v \over f_{\rm C}},
\label{eq:linear_dx}\\
v_n(t)&\kern-0.2cm=&\kern-0.2cmv_{\rm C}+\frac{{\gamma}_{\rm out}
e^{-{\gamma}_{\rm in}t}+ {\gamma}_{\rm in} e^{{\gamma}_{\rm out}t}}
{{\gamma}_{\rm in}+{\gamma}_{\rm out}}\delta v.
\label{eq:linear_v}
\end{eqnarray}
By eliminating the time $t$, we find the equation for the trajectory
around the point C,
\begin{equation}
\kern-3mm\left[\frac{af_{\rm C}(\Delta x - \Delta x_{\rm C})
-(a-\gamma_{\rm in})(v - v_{\rm C})}{\gamma_{\rm in}\delta v}\right]
^{\gamma_{\rm in}}\kern-1mm
\left[\frac{(a+\gamma_{\rm out})(v - v_{\rm C})
-af_{\rm C}(\Delta x -\Delta x_{\rm C})}{\gamma_{\rm out}\delta v}
\right]^{\gamma_{\rm out}}\kern-6mm=1.
\label{eq:linear_traj}
\end{equation}
This curve shown in Fig.~\ref{fig:linearized.eps} has two asymptotes
through C with slopes $af_{\rm C}/(a-{\gamma}_{\rm in})$ and $a f_{\rm
C}/(a+{\gamma}_{\rm out})$, which correspond to the decelerating and
accelerating asymptotic trajectories, respectively.

\begin{figure}[hbt]
\epsfxsize=8cm
\centerline{
\epsfbox{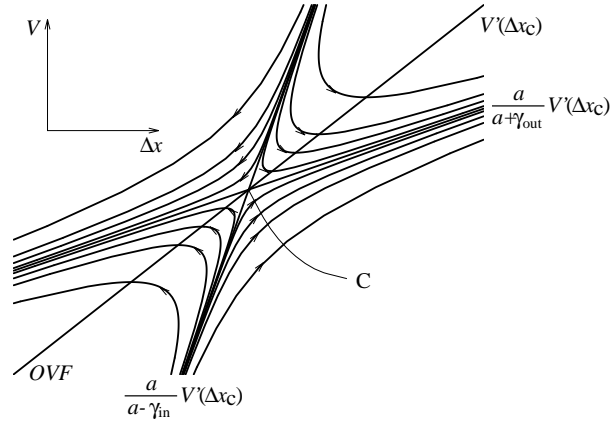}
}
\caption{
The flow diagram around the point C obtained by the linear
approximation.  Two asymptotes are asymptotic trajectories.}
\label{fig:flow.eps}
\end{figure}
Two asymptotes of (\ref{eq:linear_traj}) divide the $\Delta x$-$v$
plane into four areas.  In Fig.~\ref{fig:flow.eps}, curves are shown
for solutions to Eq.~(\ref{eq:linear_eq}) with various initial
conditions: the point C is a saddle point.  The linear analysis
applies to the point F as well, so curves in the left-lower region
would describe the behavior of vehicles close to a free region.

The condition, $(v(t)-v_{\rm C})/v_{\rm C} << 1$, helps us to evaluate
$t_{\rm C}$, the time interval a vehicle would spend around the point
C;
\begin{equation}
t^2 << \frac{2}{{\gamma}_{\rm in}{\gamma}_{\rm out}} \frac{v_{\rm
C}}{\delta v} \sim t_{\rm C}^2.
\end{equation}
The time $t_{\rm C}$ is related to the length of a congested region
$L_{\rm C}$ and the number of vehicles in this region $N_{\rm C}$ as
follows,
\begin{equation}
N_{\rm C}=\frac{t_{\rm C}}T,\qquad
L_{\rm C}=N_{\rm C}\Delta x_{\rm C}=\Delta x_{\rm C}\frac{t_{\rm C}}T
\end{equation}
Therefore the size of a congested region is larger for smaller $\delta
v$.

\section{Summary and Discussion}
\label{sec:Discussion}
We have investigated the repetitive pattern formation observed in a
computer simulation from an analytical point of view.  The Rondo
approach was proposed to describe the repetitive pattern.  In addition
to $a$ and $V(\Delta x)$, the difference-differential equation for the
Rondo function $F(t)$ contains two macroscopic parameters, $T$ and
$v_B$, which specify the motion of a global pattern.  In this paper we
mainly paid our attention to Rondo functions for asymptotic
trajectories.  The Rondo equation was solved for three simple models
with piece-wise linear OV-functions.  In order to determine the
$a$-dependence of $T$ and $v_B$, we gave analytic expressions for Rondo
functions.  We would like to emphasize that the concept of asymptotic
trajectory plays a key role to determine the $a$-dependence of $T$ and
$v_B$.  As a first step to understand more realistic situations, we have
studied some trajectories around cusps.

In the following we discuss three questions related to the Rondo
approach: (1) extension to OV-models with asymmetric OV-functions; (2)
more on realistic trajectories; (3) possibility to find the Rondo
function {\it forward} in time.

Studying more realistic models, we may encounter an asymmetric
OV-function. We explain how our procedure developed in this paper may be
extended to such situations. Even with an asymmetric OV-function, we may
define the concept of accelerating and decelerating asymptotic
trajectories.  When the symmetry is absent, accelerating and
decelerating asymptotic trajectories are not related each other and must
be found independently.  The condition that both of them share the same
end points C and F will determine the $a$-dependence of $T$.

Even though we have mainly studied asymptotic trajectories, the Rondo
equation itself must be applicable for any repetitive motion of
vehicles.  On the other hand, as we have observed for piece-wise linear
models, the difference between asymptotic trajectories and the results
of computer simulations are very small.  Therefore whether we would like
to obtain a realistic trajectory out of the Rondo approach or not is
very much dependent of our purpose.

The Rondo equation gives us a functional relation which may be written
as,
\begin{equation}
F(t+T)= {\cal P}[F(t),{\dot F}(t),{\ddot F}(t); a, v_B, T],
\label{eq:functional}
\end{equation}
with three parameters $a$, $v_B$ and $T$.  When we know the function
$F(t)$ for the time interval $t \in [t_0, t_0+T]$, there are two ways to
use the above equation: (1) substituting this on the r.h.s., we find
$F(t)$ for $t \in [t_0+T, t_0+2T]$; (2) the same information may be used
on the l.h.s. to have a differential equation for $F(t)$ on the interval
$t \in [t_0-T, t_0]$.  Although the former sounds much easier, we have
been able to use the Rondo equation only in the latter manner.  Here we
would explain the reason why it was so.

Now let us consider, for concrete, a decelerating asymptotic trajectory
in the double slope model.  In order to use Eq.~(\ref{eq:functional}) in
the approach (1), we need the Rondo function describing a part of the
asymptotic trajectory coming out of F; the rest of the asymptotic
trajectory may be obtained just by differentiating the initial function
repeatedly.  So the initial Rondo function is of vital importance.

Since the OV-function is symmetric, the transcendental equation
(\ref{eq:gamma}) may be used to find the initial Rondo function. Let us
remember how solutions are distributed.  There are only two real
solutions, $-\gamma_{\rm in} < 0$ and $\gamma_{\rm out} > 0$, and
infinitely many complex solutions whose real parts are larger than
$\gamma_{\rm out}$.  The initial Rondo function is a linear combination
of infinitely many exponential functions with ${\rm Re}(\gamma) >
0$. Therefore it contains infinitely many coefficients, which must be
determined to give an asymptotic trajectory with properties described in
section 3.  To find an asymptotic trajectory in this way, we probably
need some new techniques.

In this paper we have considered the repetitive pattern in traffic flow.
Such repetitive structure is also observed in various phenomena, and we
believe that our approach may be helpful to understand them.

\vspace*{1cm}
\appendix
\noindent
{\Large\bf Appendix}
\section{Asymptotic trajectory is symmetric}
\label{ap-sec:symmetric}

In this paper we have used the following property of an asymptotic
trajectory: $(\Delta x_{\rm C},v_{\rm C})$ and $(\Delta x_{\rm
F},v_{\rm F})$ are at symmetric positions for OV-function symmetric
around a point S.  Here we would like to give a proof of
the above claim for an OV-function symmetric with respect to the point
S.

Suppose an OV-function function and a sensitivity are given.  Our
Rondo equation contains two parameters $T$ and $v_B$;
\begin{equation}
{1\over a}\ddot F(t) + \dot F(t) = V(F(t+T) -F(t) + v_BT).
\end{equation}
If we could find a solution $F(t)$ for the equation, this means that,
for the OV-function and the sensitivity $a$, corresponding pattern
with the delay $T$ and the backward velocity $v_B$ may be realized.

The OV-function is taken to be an odd function around the point
S$(\Delta x_{\rm S}, v_{\rm S})$.  This assumption is expressed with
an odd function $W(-x)=-W(x)$ as follows,
\begin{equation}
V(\Delta x)=v_{\rm S} + W(\Delta x - \Delta x_{\rm S}).
\end{equation}
By putting this form to the Rondo equation, it now looks like,
\begin{equation}
{1\over a}\ddot F(t) + \dot F(t) = v_{\rm S} + W(F(t+T) -F(t) + v_BT -
\Delta x_{\rm S}).
\end{equation}

We assume that a solution to the Rondo equation has been found.  It
gives a trajectory on the $\Delta x$-$v$ plane, whose coordinate we
denote as $(\Delta x_1(t), v_1(t))$.  They are expressed with the
function $F(t)$ as follows,
\begin{equation}
v_1(t)={\dot F}(t),\qquad\Delta x_1(t)=F(t+T) -F(t) + v_BT.
\end{equation}

By using the fact that $W$ is odd, it is easily shown that $(\Delta
x_2(t), v_2(t))$ given below satisfies the Rondo equation as well.
\begin{equation}
v_2(t)={\dot {\bar F}}(t),
\qquad\Delta x_2(t)={\bar F}(t+T)-{\bar F}(t)+{v^\prime}_BT,
\end{equation}
where ${\bar F}(t) \equiv 2 v_{\rm S} t - F(t)$ and
${v^\prime}_BT \equiv - 2 v_{\rm S} T - v_B T + 2 \Delta x_{\rm S}$.

It is also easy to see that $(\Delta x_1(t), v_1(t))$ and $(\Delta
x_2(t), v_2(t))$ are symmetric with respect to the point S.
Therefore if the former defines a trajectory from a free to a
congested region, the latter defines that for opposite direction.

Here we emphasize that two trajectories have different backward
velocities but with the same delay time $T$.

In computer simulations, we observe that a pattern of a congested flow
is characterized with two parameters $T$ and $v_B$; both regions,
connecting free to congested or congested to free, moves with the same
backward velocity $v_B$. So two trajectories connecting free and
congested regions must have the same parameters.  This must be also
true for an asymptotic trajectory.  Therefore ${v^\prime}_B$ must be
equal to $v_B$ itself.  This implies that two trajectories discussed
above form a closed trajectory.  Thus we may conclude the following:
1) solutions expressed by $F(t)$ and ${\bar F}(t)$ satisfy the Rondo
equation with the same parameters $T$ and $v_B$ ; 2) the two points on
the OV-function connected by trajectories are symmetric with respect
to the point S; 3) the straight line through the two points
includes the point S.

\section{General formula for Step-by-Step method}
\label{ap-sec:General formula}
In the following we will give a general formula for the second order
linear differential equation,
\begin{eqnarray}
&&\left(\frac1{af}\frac{d^2}{dt^2}+\frac1{f}\frac{d}{dt}+1\right)G_k(t)
=G_{k-1}(t)\cr
&&\qquad\qquad G_k(0)=0,\quad {\dot G}_k(0)=0 \qquad (k\ge 1)
\label{eq:ap.Gk}\\
&&\left(\frac1{af}\frac{d^2}{dt^2}+\frac1{f}\frac{d}{dt}+1\right)G_0(t)
=G_0(t)\cr
&&\qquad\qquad G_0(0)=0,\quad {\dot G}_0(0)=-a\delta
\label{eq:ap.G0}
\end{eqnarray}

The solution for the Eq.~(\ref{eq:ap.G0}) is
\begin{equation}
G_0(t)=\delta\left(e^{-at}-1\right)
\end{equation}
In terms of the function $g_k(\theta)$ defined as follows,
\begin{equation}
G_{k+1}(t)=G_k(t)+\frac{a\delta}\omega\left(\frac{af}{\omega^2}
\right)^ke^{-\frac a2t}g_k(\omega t)\qquad (k\ge 0)
\label{eq:ap.def-gk}
\end{equation}
the initial conditions at $t=0$ are expressed as
\begin{eqnarray}
g_k(0)=0,&&g'_k(0)=0,\qquad (k\ge 1)\cr
g_0(0)=0,&&g'_0(0)=1.
\label{eq:ap.gk.init}
\end{eqnarray}
{}From the Eqs.~(\ref{eq:ap.Gk}), (\ref{eq:ap.G0}) and
$\omega^2=af-a^2/4$, the equation to determine $g_k(\theta)$ is
\begin{eqnarray}
&&g''_k(\theta)+g_k(\theta)=g_{k-1}(\theta)\qquad (k\ge 1)\cr
&&g''_0(\theta)+g_0(\theta)=0
\label{eq:ap.gk}
\end{eqnarray}
The initial value problem with (\ref{eq:ap.gk.init}) and
(\ref{eq:ap.gk}) may be solved by the functions
\begin{equation}
g_k(\theta)=\frac1{2^{2k}k!}\sum^k_{m=0}\frac{(2k-m)!}{(k-m)!\, m!}
(-1)^{\left[\frac{m+1}2\right]}(2\theta)^m
\left(\begin{array}{c}\sin\theta\\\cos\theta\end{array}\right)_m
\label{eq:ap.sol-gk}
\end{equation}
Here
\begin{equation}
\left(\begin{array}{c}\sin\theta\\\cos\theta\end{array}\right)_m=
\left\{\begin{array}{l}\sin\theta\qquad(m:{\rm even})\\
\cos\theta\qquad(m:{\rm odd})\end{array}\right.
\end{equation}
and $\left[r\right]$ is the maximum integer which does not exceed $r$.
We give functions for $k=0,1,2,3$ explicitly.
\begin{eqnarray}
g_0(\theta)&\kern-0.2cm=&\kern-0.2cm\sin\theta,\cr
g_1(\theta)&\kern-0.2cm=&\kern-0.2cm
\frac12\left(\sin\theta-\theta\cos\theta\right),\cr
g_2(\theta)&\kern-0.2cm=&\kern-0.2cm\frac18
\left(3\sin\theta-3\theta\cos\theta-\theta^2\sin\theta\right),\cr
g_3(\theta)&\kern-0.2cm=&\kern-0.2cm
\frac1{48}\left(15\sin\theta-15\theta\cos\theta
-6\theta^2\sin\theta+\theta^3\cos\theta\right).
\end{eqnarray}

In this article, we also use another series of solutions $h_k(t)$ of
equations (\ref{eq:ap.gk}) with initial condition;
\begin{eqnarray}
h_k(0)=0,&&h'_k(0)=0,\qquad (k\ge 1)\cr
h_0(0)=1,&&h'_0(0)=0.
\label{eq:ap.hk.init}
\end{eqnarray}
It is easily shown that $h_k(\theta)$ is given by
\begin{equation}
h_k(\theta)=g'_k(\theta)\equiv\frac{\theta}{2k}g_{k-1}(\theta),
\label{eq:ap.sol-hk}
\end{equation}
where the second equality is valid only for $k\ge1$. We also give
$h_k(\theta)$ for $k=0,1,2,3$ explicitly.
\begin{eqnarray}
h_0(\theta)&\kern-0.2cm=&\kern-0.2cm\cos\theta,\cr
h_1(\theta)&\kern-0.2cm=&\kern-0.2cm\frac12\theta\sin\theta,\cr
h_2(\theta)&\kern-0.2cm=&\kern-0.2cm
\frac18\left(\theta\sin\theta-\theta^2\cos\theta\right),\cr
h_3(\theta)&\kern-0.2cm=&\kern-0.2cm\frac1{48}
\left(3\theta\sin\theta-3\theta^2\cos\theta-\theta^3\sin\theta\right),
\end{eqnarray}


\end{document}